\documentclass[aps,preprint,nofootinbib,floatfix]{revtex4-1}
\usepackage{graphicx}
\usepackage{amsmath,amssymb}
\usepackage{url}
\usepackage{color,epstopdf}
\newcommand{\lsim}{\mathrel{\mathop{\kern 0pt \rlap
  {\raise.2ex\hbox{$<$}}}
  \lower.9ex\hbox{\kern-.190em $\sim$}}}
\newcommand{\gsim}{\mathrel{\mathop{\kern 0pt \rlap
  {\raise.2ex\hbox{$>$}}}
  \lower.9ex\hbox{\kern-.190em $\sim$}}}

\begin{document}
\title{Mimicking the Standard Model Higgs Boson in UMSSM\\}
\author{Chun-Fu Chang$^{1}$, Kingman Cheung$^{2,1}$, Yi-Chuen Lin$^{1}$, 
and Tzu-Chiang Yuan$^3$}

\affiliation{
$^1$Department of Physics, National Tsing Hua University, 
Hsinchu 300, Taiwan\\
$^2$Division of Quantum Phases \& Devices, School of Physics, 
Konkuk University, Seoul 143-701, Republic of Korea \\
$^3$Institute of Physics, Academia Sinica, Nankang, Taipei 11529, Taiwan
}

\date{\today} 

\begin{abstract}
  Motivated by the recent results in the standard model (SM) Higgs
  boson search at the Large Hadron Collider (LHC) we investigate the
  SM-like CP-even Higgs boson of the $U(1)'$-extended minimal
  supersymmetric standard model (UMSSM) and its branching ratio into
  the $b\bar b$, $W W^*$, and $\tilde{\chi}^0_1 \tilde{\chi}^0_1$
  modes.  In the Summer 2011, a $2\sigma$ excess was reported in the
  channel $H \to W W^* \to \ell^+ \nu \ell^- \bar \nu$ around $130 -
  140$ GeV range.  Later on in December 2011 announcements were made
  that an excess was seen in the $124-126$ GeV range, while the SM Higgs
  boson above 131 GeV up to about 600 GeV is ruled out.  We examine
  two scenarios of these mass ranges: (i) $130 \,{\rm GeV} <
  M_{h_{\rm SM-like}} < 141\,{\rm GeV}$ and show that the Higgs boson
  can decay into invisible neutralinos to evade the SM bound; and (ii)
  $120 \,{\rm GeV} < M_{h_{\rm SM-like}} < 130\,{\rm GeV}$ and show
  that the Higgs boson can avoid decaying into neutralinos and thus
  gives enhanced rates into visible particles.  We use the $\eta$
  model of $E_6$ with TeV scale supersymmetry to illustrate the idea
  by scanning the parameter space to realize these two different
  scenarios.
\end{abstract}
\maketitle

\section{Introduction}

The excitement of particle physics in the year 2011 was the hunt for
the Higgs boson, the Higgs boson of any model, in particular that of
the standard model (SM) at the Large Hadron Collider (LHC). 
The recent data in the Summer 2011 \cite{recent}, 
showed an approximately $2\sigma$ excess in the channel 
$W W^* \to \ell^+ \nu \ell^- \bar \nu$ ($\ell =e,\mu$) above the
expected SM backgrounds.  The excess is consistent with a Higgs boson
of mass about 140 GeV but with a somewhat smaller production rate
of $H \to W W^* \to \ell^+ \nu \ell^- \bar \nu$ than the SM one.
However, in December 2011 both ATLAS \cite{126-atlas} and 
CMS \cite{126-cms} announced
possible hints of excess in $\gamma\gamma$, $WW$, and $ZZ$ channels
that are consistent with a SM-like Higgs boson in the mass range of 
$124-126$ GeV; and at the same time rule out a SM Higgs boson above 
131 GeV up to about 600 GeV.  
Except for the $\gamma\gamma$ channel almost all channels are slightly 
suppressed relative to the SM cross sections at around $124-126$ GeV.
Note that these results consist of large errors. 
In this work, we consider two mass ranges, $120-130$ and $130-141$ GeV,
for the SM-like Higgs boson of the $U(1)'$-extended minimal supersymmetric
standard model (UMSSM).  We will entertain these two ranges in the 
supersymmetry (SUSY) 
framework, because the current data are still premature to 
definitely confirm a Higgs boson, not to mention its mass.   

Supersymmetric models in general predicts a light Higgs boson, mostly
below about 150 GeV.  In particular, the minimal supersymmetric 
standard model (MSSM) predicts a light Higgs boson with $M_{h} \alt 130$ GeV.
Thus, a Higgs boson heavier than 130 GeV significantly constrains the
parameter space of the MSSM, forcing the sfermions masses to exceed the
TeV range, and consequently SUSY loses somewhat of its appeal.
It is then more natural to consider extensions to the MSSM
if the light Higgs boson is heavier than 130 GeV, and to hide this Higgs
boson by suppressing its branching ratio into visible modes.
It is well known that by adding singlet Higgs field can easily raise
the Higgs boson mass.
Recent attempts to raise the Higgs boson mass in SUSY frameworks can be
found in Refs.~\cite{kolda,baer,tj,naka,ma1,endo,asano,elsa},
and attempts to hide such a light Higgs boson heavier than 130 GeV 
in the current data can be found in 
Refs.~\cite{he,raid,ell,carlo,mam,ma,low,eng,leb,foot,posp,weihs,cui,csaki,bai,godbole,krib,djou,eng2,ko,ltw}.

On the other hand, if the SM-like Higgs boson falls in the mass range 
of $124-126$ GeV and future data may further support that, this Higgs boson
should decay into visible particles, almost in the same pattern as
the SM Higgs boson, though the current data \cite{126-atlas,126-cms}
showed a slightly enhancement to
the $\gamma\gamma$ mode while slightly suppression to the $WW$, $ZZ$, and
$b\bar b$ modes.  Recent attempts interpreting the $124-126$ GeV Higgs boson
in SUSY framework can be found in Refs.~\cite{mssm}. 
In order to give a $124-126$ GeV Higgs boson within MSSM, the stop
sector must consist of a very heavy stop, a large mixing, and a relatively
light stop, which has an interesting implication to collider phenomenology.
However, within the MSSM it is rather difficult to enhance the $\gamma\gamma$
production rate but easier in some 
other extensions like the Randall-Sundrum scenario \cite{126-RS} 
and others \cite{arhrib}.

It is therefore
timely to investigate an extension of MSSM, which involves an
extra $U(1)'$ symmetry and a Higgs singlet superfield $S$.
The scalar component of the Higgs singlet
superfield develops a vacuum expectation value (VEV), which 
breaks the $U(1)'$ symmetry and gives a mass to the $U(1)'$ gauge boson,
denoted by $Z'$.  At the same time, the VEV together with the Yukawa
coupling can form an effective $\mu_{\rm eff}$ parameter from the term
$\lambda \langle S \rangle H_u H_d = \mu_{\rm eff} H_u H_d$ in the
superpotential, thus solving the $\mu$ problem of MSSM (the same as
the NMSSM \cite{ellwanger}).

Existence of extra neutral gauge bosons had been predicted in many extensions
of the SM \cite{paul}.  String-inspired models and
grand-unification (GUT) models usually contain a number of extra $U(1)$
symmetries, beyond the hypercharge $U(1)_Y$ of the SM.  The exceptional group
$E_6$ is one of the famous examples of this type.
Phenomenologically, the most interesting option
is the breaking of these $U(1)$'s at around TeV scales, giving rise to
an extra neutral gauge boson observable 
at the Tevatron and the LHC.
In a previous work \cite{ours}, 
we considered a scenario of $U(1)'$ symmetry breaking at
around TeV scale by the VEV of a Higgs singlet superfield in the
context of weak-scale supersymmetry.  The $Z'$ boson obtains a mass from
the breaking of this $U(1)'$ symmetry and proportional to the VEVs.
Such a $Z'$ can decay into the SUSY particles such as neutralinos,
charginos, and sleptons, in addition to the SM particles.  Thus, the
current mass limits are reduced by a substantial amount and so is the
sensitivity reach at the LHC \cite{ours,kang}.

In this work, we turn our focus to the Higgs sector
in UMSSM, which consists
of 3 CP-even Higgs bosons, 1 CP-odd Higgs boson, and one pair of 
charged Higgs bosons. Because of the extra singlet Higgs superfield
the mass of the lightest Higgs boson is raised by a substantial 
amount, easily to be above 130 GeV. However, for such a heavy Higgs boson,
we have to hide it under the current data because the SM Higgs boson is 
ruled out for above 141 GeV \cite{126-atlas,126-cms}.
In UMSSM, there is an invisible decay mode of the SM-like
Higgs boson into a pair of lightest neutralinos.  We shall show that
there are substantial parameter space that it is possible to 
hide the SM-like Higgs boson in this manner.
On the other hand, if the SM-like Higgs boson lies in the lower mass range
$120-131$ GeV, we can find the parameter space that this SM-like
Higgs boson decays in a manner similar to the SM Higgs boson, i.e.,
the decay branching ratios into $\gamma\gamma$, $WW$, $ZZ$, 
and $b\bar b$ are all similar to the SM values.

So what are the differences between the UMSSM and the other
ones such as the next-to-minimal supersymmetric standard model (NMSSM)?  
There are a number of extensions to the MSSM by
adding a Higgs singlet superfield depending on the different extra terms
of the singlet in the superpotential (such as $\kappa S^3$ of the NMSSM).  
However, these extensions often predict a light CP-odd Higgs boson
in addition to the usual MSSM-like CP-odd Higgs boson, such that the
SM-like Higgs boson decays dominantly into this light CP-odd Higgs boson.
In such a way, the branching ratios into 
$\gamma\gamma$, $WW$, $ZZ$, and $b\bar b$ diminish to negligible
values, and so cannot explain the excess seen at the LHC.  The point here
is that not all Higgs-singlet extensions to the MSSM can account for the
excess at the LHC, although most of them can raise the Higgs boson to 
the desirable value.  The UMSSM, on the other hand,
 only has one CP-odd Higgs boson, which is MSSM-like.

We organize the paper as follows. In the next section, we describe the
model briefly and work out the mass matrix of the CP-even Higgs bosons.  
In Sec. III, we list the formulas for the couplings of the CP-even Higgs 
bosons that are most relevant to our study.
In Sec. IV, we search for the parameter space in the model that can have a
SM-like Higgs boson in the two mass ranges of (i) $130-141$ GeV 
and (ii) $120-130$ GeV, and show the branching ratios into
$WW$, $b\bar b$, and $\tilde{\chi}^0_1 \tilde{\chi}^0_1$.
We discuss and conclude in Sec. V.

\section{UMSSM}

For illustration we use the popular grand unified models of 
$E_6$, which are anomaly-free.
Two most studied $U(1)$ subgroups in the symmetry breaking chain 
of $E_6$ are
\[
  E_6 \to SO(10) \times U(1)_\psi\,, \qquad 
 SO(10) \to SU(5) \times U(1)_\chi  \;.
\]
In $E_6$ each family of the left-handed fermions is promoted to a 
fundamental $\mathbf{27}$-plet, which decomposes under 
$E_6 \to SO(10) \to SU(5)$ as 
\[
\mathbf{27} \to \mathbf{16} + \mathbf{10} + \mathbf{1} \to
  ( \mathbf{10} + \mathbf{5^*} + \mathbf{1} ) + (\mathbf{5} +
  \mathbf{5^*} ) + \mathbf{1} \;.
\]
We require one of the $U(1)$'s remains unbroken until around TeV scale, and then
broken to give masses to the $Z'$ boson and its superpartner $Z'$-ino.  
Each $\mathbf{27}$ contains the SM fermions, 
two additional singlets 
$\nu^c$ (conjugate of the right-handed neutrino) and $S$, 
a $D$ and $D^c$ pair ($D$ is the exotic color-triplet quark 
with charge $-1/3$ and $D^c$
is the conjugate), and a pair of color-singlet SU(2)-doublets
$H_u$ and $H_d$ 
with hypercharge $Y_{H_u, H_d} = \pm 1/2$.  In the supersymmetric
version of $E_6$, the scalar components of one $H_{u,d}$ pair 
(out of 3 if there are 3 families)
can be used as the two Higgs doublets $H_{u,d}$ of the MSSM. 
The chiral charges $U(1)_\chi$ and $U(1)_\psi$
for each member of the $\mathbf{27}$
are listed, respectively, in the third and fourth columns in Table~\ref{E6}.
In general, the two $U(1)_\psi$ and $U(1)_\chi$ can mix to form
\begin{equation}
  Q'(\theta_{E_6} ) = \cos \theta_{E_6} Q'_\chi + \sin \theta_{E_6} Q'_\psi \;,
\end{equation}
where $0 \le \theta_{E_6} < \pi$ is the mixing angle.  A commonly studied
model is the $Z'_\eta$ model with 
\begin{equation}
\label{eta-model}
 Q'_\eta = \sqrt{ \frac{3}{8} } Q'_\chi - \sqrt{ \frac{5}{8} } Q'_\psi \;,
\end{equation}
which has $\theta_{E_6} = \pi - \tan^{-1} \sqrt{5/3} \sim 0.71 \pi$. 
There are also the inert model with
$Q'_I = -Q'(\theta_{E_6} = \tan^{-1} \sqrt{3/5} \sim 0.21 \pi)$,
the neutral $N$ model with 
$\theta_{E_6} = \tan^{-1} \sqrt{15} \sim 0.42 \pi$, and
the secluded sector model with 
$\theta_{E_6} = \tan^{-1} \sqrt{15}/9 \sim 0.13 \pi$.
The chiral charges for each member of the $\mathbf{27}$
are also listed in the last four columns in Table~\ref{E6} for 
these four variations of $Z'$ models within $E_6$.
Here we take the assumption that all the exotic particles, other than the 
particle contents of the MSSM, are very heavy and well beyond the reaches
of all current and planned colliders. For an excellent review of $Z'$ models,
see Ref. \cite{paul}.

\begin{table}[tbh!]
\caption{\small \label{E6}
The chiral charges of the left-handed fermions for various $Z'$ models 
arised in $E_6$ \cite{paul}. Note that 
$Q'_{f_R} = - Q'(f^c)$ since all the right-handed SM fermions are necessarily
charge-conjugated to convert into left-handed fields in order to put them into
the irreducible representation of $\mathbf{27}$ of $E_6$.
}
\smallskip
\begin{ruledtabular}
\label{table-models}
\begin{tabular}{cccccccc}
$SO(16)$ & $SU(5)$ & $2\sqrt{10} Q'_\chi$  & $2\sqrt{6}Q'_\psi$ & 
$2\sqrt{15}Q'_\eta$ & $2Q'_I$ & $2\sqrt{10}Q'_N$  & $2\sqrt{15}Q'_{\rm sec}$ \\ 
\hline
$\mathbf{16} $ & $\mathbf{10}(u,d, u^c, e^c)$ & $-1$ & $1$ & $-2$ & $0$ &
                  $1$ & $-1/2$ \\
                & $\mathbf{5^*}(d^c,\nu, e^-)$ & $3$ & $1$ & $1$ & $-1$ &
                  $2$ & $4$ \\
                & $\nu^c$ & $-5$ & $1$ & $-5$ & $1$ & $0$ & $-5$ \\
\hline
$\mathbf{10} $ & $\mathbf{5}(D, H_u)$ & $2$ & $-2$ & $4$ & $0$ &
                  $-2$ & $1$ \\
                & $\mathbf{5^*}(D^c, H_d)$ & $-2$ & $-2$ & $1$ & $1$ &
                  $-3$ & $-7/2$ \\
\hline
$\mathbf{1} $ & $\mathbf{1} S $ & $0$ & $4$ & $-5$ & $-1$ &   $5$ & $5/2$ 
\end{tabular}
\end{ruledtabular}
\end{table}

The effective superpotential $W_{\rm eff}$ involving the matter and Higgs superfields in UMSSM can be written as 
\begin{equation}
\label{sp}
W_{\rm eff} = \epsilon_{ab} \left [ y^u_{ij} Q^a_j H_u^b U^c_i 
  - y^d_{ij} Q^a_j H_d^b D^c_i 
  -  y^l_{ij} L^a_j H_d^b E^c_i 
  +  h_s S  H_u^a H_d^b \right ] \;,
\end{equation}
where $\epsilon_{12}= - \,\epsilon_{21} =1$, $i,j$ are family indices,
and $y^u$ and $y^d$ represent the Yukawa matrices for the 
up-type and down-type quarks respectively.
Here $Q, L, U^c, D^c, E^c, H_u$, and $H_d$ denote the MSSM superfields for the 
quark doublet, lepton doublet,
up-type quark singlet, down-type quark singlet, lepton singlet,
up-type Higgs doublet, and down-type Higgs doublet respectively,
and the $S$ is the singlet superfield. 
Note that we have assumed other exotic fermions are so heavy 
that they have been integrated out and do not enter
into the above effective superpotential.
The $U(1)'$ charges of the fields
$H_u, H_d,$ and $S$ are related by $Q'_{H_u} + Q'_{H_d} + Q'_S = 0$ such 
that $S H_u H_d$ is the only term allowed by the $U(1)'$ symmetry 
beyond the MSSM. Once the singlet scalar field $S$ develops a 
VEV, it generates an effective $\mu$ parameter: $\mu_{\rm eff} = 
h_s \langle S \rangle$.  The case is very similar to NMSSM, except we
do not have the cubic term $S^3$ 
since it is forbidden by the $U(1)'$ symmetry.

The singlet superfield will give rise to a singlet scalar boson and a
singlino. The real part of the scalar boson will mix with the real
part of $H_u^0$ and $H_d^0$ to form 3 CP-even Higgs boson.  The
imaginary part of the singlet scalar will be eaten in the process
of $U(1)'$ symmetry breaking
 and becomes the longitudinal part of the $Z'$
boson.  The singlino, together with the $Z'$-ino, will mix with the
neutral gauginos and neutral Higgsinos to form 6 neutralinos.
Studies of various singlet-extensions of the MSSM can be found in 
Refs.~\cite{vernon-n,vernon-h}.
At or below TeV scale the particle content is almost the same as the
MSSM, except that it has 3 CP-even Higgs boson, 1 CP-odd Higgs boson,
and a pair of charged Higgs boson in the Higgs sector, and also 
a $Z'$ boson and 2 extra neutralinos  (coming from the $Z'$-ino and
the singlino.)

The gauge interactions involving the fermionic and scalar 
components, denoted generically by $\psi$ and $\phi$ respectively,  
of each superfield are
\begin{equation}
  \label{eq1}
 {\cal L} = \frac{1}{2} \bar  \psi_i \, i \gamma^\mu D_\mu \, \psi_i 
   + ( D^\mu \phi_i)^\dagger \, (D_\mu \phi_i) \;,
\end{equation}
where $\psi_i$ and $\phi_i$ denote the Majorana fermionic and bosonic
components of the superfield, respectively.
The covariant derivative of $\phi_i$ is given by 
\begin{equation}
\label{eq2}
  D_\mu \phi_i = 
\left[\partial_\mu + i e Q A_\mu + i \frac{g }{\sqrt{2}} (\tau^+ W^+_\mu
  +\tau^- W^-_\mu ) + i g_1 ( T_{3L} - Q x_{\rm w} ) Z_\mu 
        + i g_2  Z'_\mu  Q' \right ]\, \phi_i \;.
\end{equation}
Here $e$ is the electromagnetic coupling constant, $Q$ is the electric 
charge, $g$ is the $SU(2)_L$ coupling,
$\tau^\pm$ are the rising and lowering operators on weak doublets,
$T_{3L}$ is the third component of the weak isospin,
and $Q'$ is the chiral charges of the $U(1)'$ associated with
the $Z'$ boson.  The interactions of $Z'$ with all MSSM fields
go through Eqs.~(\ref{eq1}) and (\ref{eq2}).
The chiral charges of various $Z'$ models are listed in Tables~\ref{E6}.
The coupling constant $g_1$ in Eq. (\ref{eq2}) is the 
SM coupling $g/\cos\theta_{\rm w}$, while
in grand unified theories (GUT) $g_2$ is related to $g_1$ by
\begin{equation}
\label{g2g1}
\frac{g_2}{g_1} = \left(\frac{5}{3}\, x_{\rm w} \lambda\right)^{1/2} \simeq
0.62\lambda^{1/2} \,,
\end{equation}
where $x_{\rm w}=\sin^2\theta_{\rm w}$ and $\theta_{\rm w}$ is the weak mixing
angle.
The factor $\lambda$ depends on the symmetry breaking pattern and the fermion
sector of the theory, which is usually of order unity.

The Higgs doublet and singlet fields are 
\begin{equation}
H_d = \left( \begin{array}{c}
               H_d^0 \\
               H_d^- \end{array}  \right ) \;\; , \qquad 
H_u = \left( \begin{array}{c}
               H_u^+ \\
               H_u^0 \end{array}  \right )  \;\; \qquad {\rm and} \qquad 
S  \;.
\end{equation}
The scalar interactions are obtained by calculating the 
$F$- and $D$-terms of the superpotential, and by including the 
soft-SUSY-breaking terms.  The terms involving the neutral components
of the Higgs fields are 
\begin{equation}
V_{H} = V_F + V_D + V_{\rm soft} \;,
\end{equation}
with
\begin{eqnarray}
V_F &=& | - h_s H_u^0 H_d^0 |^2 + |h_s S|^2 ( |H_u^0|^2 + |H_d^0|^2 ) \;, \\
V_D &=& \frac{g_1^2}{8} \left( | H_u^0 |^2  - |H_d^0|^2 \right )^2 
   + \frac{g_2^2}{2}  \left( Q'_{H_u}| H_u^0 |^2  +Q'_{H_d} |H_d^0|^2 
         + Q'_S |S|^2 \right )^2  \;, \\
V_{\rm soft} &=& M^2_{H_u} | H_u^0 |^2 + M^2_{H_d} | H_d^0 |^2 +M_S^2 |S|^2
            + ( - h_s A_s S H_u^0 H_d^0 + {\rm h.c.} ) \;.
\end{eqnarray}
The minimization conditions of $\partial V_{H} / \partial H_u^0 = 0$, 
$\partial V_{H} / \partial H_d^0 = 0$, and 
$\partial V_{H} / \partial S = 0$ at the vacuum 
give the following tadpole conditions:
\begin{eqnarray}
M_{H_u}^2 &=&  - \frac{v_d^2}{2} 
              \left( h_s^2 - \frac{g_1^2}{4} + g_2^2 Q'_{H_d} Q'_{H_u} \right )
   -  \frac{v_u^2}{2} 
           \left( \frac{g_1^2}{4} + g_2^2 Q'^2_{H_u} \right ) 
   - \frac{v_s^2}{2} 
            \left( h_s^2 + g_2^2 Q'_{S} Q'_{H_u} \right ) \nonumber \\
    & + & \frac{h_s A_s}{\sqrt{2}} \frac{v_s v_d}{v_u} \;, \\
M_{H_d}^2 &=&  - \frac{v_d^2}{2} 
              \left( \frac{g_1^2}{4} + g_2^2 Q'^2_{H_d} \right )
   -  \frac{v_u^2}{2} 
           \left( h_s^2 - \frac{g_1^2}{4} + g_2^2 Q'_{H_u} Q'_{H_d} \right ) 
   - \frac{v_s^2}{2} 
            \left( h_s^2 + g_2^2 Q'_{S} Q'_{H_d} \right ) \nonumber \\
    & + & \frac{h_s A_s}{\sqrt{2}} \frac{v_s v_u}{v_d} \;, \\
M_{S}^2 &=&  - \frac{v_d^2}{2} 
              \left( h_s^2 + g_2^2 Q'_{H_d} Q'_{S} \right )
   -  \frac{v_u^2}{2} 
           \left( h_s^2  + g_2^2 Q'_{H_u} Q'_S \right ) 
   - \frac{v_s^2}{2} g_2^2 Q'^2_{S} 
    + \frac{h_s A_s}{\sqrt{2}} \frac{v_u v_d}{v_s} \;.
\end{eqnarray}
where $\langle H_u^0 \rangle = v_u/\sqrt{2}$, 
$\langle H_d^0 \rangle = v_d/\sqrt{2}$, and
$\langle S \rangle = v_s/\sqrt{2}$ are the VEVs. 
The two VEVs $v_u$ and $v_d$ satisfy
$v^2 \equiv v_u^2 + v_d^2 = (246\;{\rm GeV})^2$ and the ratio 
$\tan\beta \equiv v_u/v_d$ is commonly defined in the literature.
Now we can expand the Higgs fields as 
\begin{eqnarray}
H_d^0 &=& \frac{1}{\sqrt{2}} \, \left( v_d + \phi_d + i \chi_d \right)
         \,,\nonumber\\
H_u^0 &=& \frac{1}{\sqrt{2}} \, \left( v_u + \phi_u + i \chi_u \right )
   \,,\nonumber\\
H_d^0 &=& \frac{1}{\sqrt{2}} \, \left( v_s + \phi_s + i \chi_s \right )
       \,,\nonumber
\end{eqnarray}
and substitute into $V_F, V_D$, and $V_{\rm soft}$. 
The tree level mass matrix ${\cal M}^{\rm tree}$ can be read off 
from the bilinear terms in the expansion, namely
\begin{equation}
V_{\rm mass} = \frac{1}{2} \left( \phi_d\,,\; \phi_u\,,\; \phi_s \right )\,
  {\cal M^{\rm tree} } \, \left( \begin{array}{c} 
                                 \phi_d \\
                                 \phi_u \\
                                 \phi_s  \end{array} \right ) \;,
\end{equation}
with
\begin{eqnarray}
{\cal M}^{\rm tree}_{11} &=& \left( \frac{g_1^2}{4} + g_2^2 Q'^2_{H_d} \right )\, 
        v_d^2 + \frac{h_s A_s}{\sqrt{2}} \frac{v_s v_u}{ v_d }\,, \nonumber\\
{\cal M}^{\rm tree}_{22} &=& \left( \frac{g_1^2}{4} + g_2^2 Q'^2_{H_u} \right )\, 
        v_u^2 + \frac{h_s A_s}{\sqrt{2}} \frac{v_s v_d}{ v_u }\,, \nonumber\\
{\cal M}^{\rm tree}_{33} &=& g_2^2 Q'^2_{S} v_S^2 
        + \frac{h_s A_s}{\sqrt{2}} \frac{v_d v_u}{ v_s }\,, \nonumber\\
{\cal M}^{\rm tree}_{12} &=& \left( h_s^2 - \frac{g_1^2}{4} + 
       g_2^2 Q'_{H_u} Q'_{H_d} \right )\,  v_d v_u 
  - \frac{h_s A_s}{\sqrt{2}} v_s = {\cal M}^{\rm tree}_{21} \,, \nonumber\\
{\cal M}^{\rm tree}_{13} &=& \left( h_s^2 + g_2^2 Q'_{H_d} Q'_{S}\right )\,v_d v_s 
  - \frac{h_s A_s}{\sqrt{2}} v_u = {\cal M}^{\rm tree}_{31} \,, \nonumber\\
{\cal M}^{\rm tree}_{23} &=& \left( h_s^2 + g_2^2 Q'_{H_u} Q'_{S}\right )\,v_u v_s 
  - \frac{h_s A_s}{\sqrt{2}} v_d = {\cal M}^{\rm tree}_{32} \,. \nonumber 
\end{eqnarray}
It is well-known that the lightest CP-even Higgs boson mass receives
a substantial radiative mass correction in MSSM.  The same is true here for the
UMSSM. Radiative corrections to the mass matrix 
${\cal M}^{\rm tree}$ have been given in Ref.~\cite{vernon-h}. 
We have included radiative corrections in our calculation.
The real symmetric mass matrix ${\cal M}={\cal M}^{\rm tree+radiative}$
can then be diagonalized by
an orthogonal transformation 
\begin{equation}
\left( \begin{array}{c} 
               h_1 \\
               h_2 \\
               h_3  \end{array} \right ) = {\cal O} \,
 \left( \begin{array}{c} 
                    \phi_d \\
                    \phi_u \\
                     \phi_s  \end{array} \right )  \qquad
{\rm or} \qquad
 \left( \begin{array}{c} 
                    \phi_d \\
                    \phi_u \\
                     \phi_s  \end{array} \right ) 
  = {\cal O}^T \,
\left( \begin{array}{c} 
               h_1\ \\
               h_2 \\
               h_3  \end{array} \right )  \;,
\end{equation}
such that ${\cal O}{\cal M}{\cal O}^T = {\rm diag}( m^2_{h_1},\;
m^2_{h_2}, \; m^2_{h_3} )$ in ascending order.
The mass spectra for the neutral CP-odd and the pair of charged Higgs bosons
are the same as MSSM.

\section{Couplings relevant for Higgs decays}

In this section, we present the neutral CP-even Higgs bosons couplings
with the gauge bosons, quarks and neutralinos.  Other couplings that
are not relevant to this work will be omitted.

The interactions of physical Higgs bosons $h_{1,2,3}$ with SM particles 
and other SUSY
particles can be obtained by writing down the Lagrangian in the weak eigenbasis
and then rotating the Higgs weak eigenstates as
\begin{eqnarray}
\phi_d &=& {\cal O}_{11} h_1 + {\cal O}_{21} h_2 + {\cal O}_{31} h_3\;\;,
\nonumber\\
\phi_u &=& {\cal O}_{12} h_1 + {\cal O}_{22} h_2 + {\cal O}_{32} h_3\;\;,
\label{higgs} \\
\phi_s &=& {\cal O}_{13} h_1 + {\cal O}_{23} h_2 + {\cal O}_{33} h_3\;\;  . 
\nonumber
\end{eqnarray}

\subsection{Higgs Couplings to Gauge Bosons}
The couplings of the Higgs bosons to a pair of gauge bosons come from
$(D_\mu H_u)^\dagger (D^\mu H_u) + (D_\mu H_d)^\dagger (D^\mu H_d)
+ (D_\mu S)^\dagger (D^\mu S)$:
\begin{equation}
{\cal L}_{\rm gauge} = {\cal L}_{WW} + {\cal L}_{ZZ} + {\cal L}_{Z'Z'} + {\cal L}_{Z Z'} \;\; ,
\end{equation}
where
\begin{eqnarray}
{\cal L}_{WW} &=& \frac{g^2}{4} W^+_\mu W^{-\mu}  
         \left [ v^2 + 2 v_u \phi_u  + 2 v_d \phi_d + \cdots  \right ] \;\; ,\nonumber \\
  &  =& m_W^2 W^+_\mu W^{-\mu}  + g m_W W^+_\mu W^{-\mu}
  \left[ (\sin\beta {\cal O}_{j2} +\cos\beta {\cal O}_{j1} ) h_j + \cdots \right ]  \;\; ,
      \\
{\cal L}_{ZZ} &=& \frac{g_1^2}{8} Z_\mu Z^{\mu}  
         \left [ v^2 + 2 v_u \phi_u  + 2 v_d \phi_d + \cdots \right ]  \;\; ,\nonumber \\
   & =& \frac{m_Z^2}{2} Z_\mu Z^{\mu}  + \frac{g_1}{2} m_Z Z_\mu Z^{\mu}
  \left[ (\sin\beta {\cal O}_{j2} +\cos\beta {\cal O}_{j1} ) h_j +\cdots \right ]  \;\; ,
   \\
{\cal L}_{Z'Z'} &=& g_2^2 Z'_\mu Z'^{\mu} \left[ Q'^2_{H_d} |H_d^0|^2 
   + Q'^2_{H_u} |H_u^0|^2 + Q'^2_{S} |S |^2  \right ]  \;\; ,\nonumber \\
  &=& \frac{1}{2} m_{Z'}^2 Z'_\mu Z'^{\mu} 
     + g_2^2 Z'_\mu Z'^{\mu} h_j v \left[ \sin\beta Q'^2_{H_u} {\cal O}_{j2} 
       + \cos\beta Q'^2_{H_d} {\cal O}_{j1} + \frac{v_s}{v} Q'^2_{S} {\cal O}_{j3}
       \right ]  \;\; ,\\
{\cal L}_{Z Z'} &=& 2 g_1 g_2 Z_\mu Z'^{\mu} \left[ \frac{1}{2} Q'_{H_d} |H_d^0|^2
    - \frac{1}{2} Q'_{H_u} |H_u^0|^2 \right ]  \;\; ,\nonumber \\
 &=& \frac{g_1 g_2}{2} Z_\mu Z'^{\mu} 
    \left[ Q'_{H_d} v_d^2 - Q'_{H_u} v_u^2 \right ]
+ g_1 g_2 Z_\mu Z'^{\mu} h_j v \left[ \cos\beta Q'_{H_d} {\cal O}_{j1} 
                             - \sin\beta Q'_{H_u} {\cal O}_{j2} \right ]  \; ,
\end{eqnarray}
with $m_W = \frac{g}{2} v,\; m_Z = \frac{g_1}{2} v$ 
and 
$m_{Z'} \approx {g_2} ( Q'^2_{H_u} v_u^2 +Q'^2_{H_d} v_d^2 + Q'^2_{s} v_s^2 )^{1/2}$ for small $Z-Z'$ mixing.

\subsection{Yukawa Couplings} 
Yukawa couplings are obtained by taking second order derivatives
of the effective superpotential in Eq.~(\ref{sp}). The interactions only 
go through the Higgs doublets, given by
\begin{eqnarray}
{\cal L}_{\rm Yukawa} &=& - \frac{g m_u}{2 m_W  \sin\beta}\, \bar u u \, \phi_u
  - \frac{g m_d}{2 m_W  \cos\beta}\, \bar d d \, \phi_d \; ,\nonumber \\
&=& - \frac{g m_u}{2 m_W  \sin\beta}\, {\cal O}_{j2} \bar u u \, h_j
    - \frac{g m_d}{2 m_W  \cos\beta}\, {\cal O}_{j1} \bar d d \, h_j \;,
\end{eqnarray}
Similar formulas can be written down for the SM leptons.

\subsection{Higgs Couplings to the Neutralinos}  
This is relevant when the
lightest neutralino is very light such that the Higgs boson can decay into.
The sources of neutralino masses come from soft masses of gauginos,
from the superpotential term $h_s S H_u H_d$, and from those supersymmetric 
couplings $-\sqrt{2} g_a \phi^\dagger T^a \overline{\tilde{\lambda}} \psi$ 
($\tilde{\lambda}$ is the Majorana gaugino field of a vector superfield, 
while $\phi$ and $\psi$ are the scalar and fermionic components of a 
matter chiral superfield).
The relevant terms for the masses are
\begin{eqnarray}
{\cal L}_{\rm neutralinos}^{\rm mass} &=& - \frac{1}{2} M_1 \overline{\tilde{B}} \tilde{B}
- \frac{1}{2} M_2 \overline{\tilde{W}^a} \tilde{W}^a
      - \frac{1}{2} M_{\tilde{Z}'} \overline{\tilde{Z}'} \tilde{Z}' \nonumber \\
&- & \frac{1}{2} \left[  
   - \mu_{\rm eff} \left( \overline{\tilde{h}_u^0} \tilde{h}_d^0 
               +  \overline{\tilde{h}_d^0} \tilde{h}_u^0  \right )
   -\frac{h_s}{\sqrt{2}} v_u  \left( 
         \overline{\tilde{S}} \tilde{h}_d^0 
       + \overline{\tilde{h}_d^0} \tilde{S}  \right )
   -\frac{h_s}{\sqrt{2}} v_d  \left( 
         \overline{\tilde{S}} \tilde{h}_u^0 
       + \overline{\tilde{h}_u^0} \tilde{S}  \right )  \right ]
   \nonumber \\
&-& \frac{1}{2} \biggr [ \frac{e }{2 c_w} v_u  \left(
 \overline{\tilde{B}} \tilde{h}_u^0 + \overline{\tilde{h}_u^0} \tilde{B} \right )
  - \frac{e }{2 c_w} v_d  \left(
 \overline{\tilde{B}} \tilde{h}_d^0 + \overline{\tilde{h}_d^0} \tilde{B} \right )
 \\
&&  - \frac{g}{2} v_u \left(  \overline{\tilde{W}^3} \tilde{h}_u^0 + 
                \overline{\tilde{h}_u^0} \tilde{W}^3 \right )
  + \frac{g}{2} v_d \left(  \overline{\tilde{W}^3} \tilde{h}_d^0 + 
                \overline{\tilde{h}_d^0} \tilde{W}^3 \right )     \nonumber \\
&&  + g_2 Q'_{H_u} v_u \left( 
   \overline{\tilde{Z}'} \tilde{h}_u^0 + \overline{\tilde{h}_u^0} \tilde{Z}'
       \right )
  + g_2 Q'_{H_d} v_d \left( 
   \overline{\tilde{Z}'} \tilde{h}_d^0 + \overline{\tilde{h}_d^0} \tilde{Z}'
       \right )
  + g_2 Q'_{S} v_s \left( 
   \overline{\tilde{Z}'} \tilde{S} + \overline{\tilde{S}} \tilde{Z}'
       \right ) \biggr ] \;\; . \nonumber \label{inte1} 
\end{eqnarray}
Thus the neutralino mass matrix ${\cal M}_N$ in the basis of 
$(\tilde{B},\,\tilde{W}^3,\, \tilde{h}_d^0,\,\tilde{h}_u^0,\,
\tilde{S},\,\tilde{Z}' )^T$ is given by
\begin{equation} {\cal M}_N = \left (
\begin{array}{cccc|cc}
M_1 & 0 & - \frac{e}{2c_w} v_d & \frac{e}{2 c_w} v_u & 0 & 0 \\
0  & M_2& \frac{g}{2} v_d & -\frac{g}{2} v_u & 0& 0 \\
- \frac{e}{2c_w} v_d & \frac{g}{2} v_d &0 & -\mu_{\rm eff}& -\frac{h_s}{\sqrt{2}}
  v_u & g_2 Q'_{H_d} v_d \\
 \frac{e}{2c_w} v_u & -\frac{g}{2} v_u & -\mu_{\rm eff}&0& -\frac{h_s}{\sqrt{2}}
  v_d & g_2 Q'_{H_u} v_u \\
\hline
0 & 0 & -\frac{h_s}{\sqrt{2}} v_u &-\frac{h_s}{\sqrt{2}} v_d &0&g_2 Q'_S v_s\\
0 & 0 & g_2 Q'_{H_d} v_d & g_2Q'_{H_u} v_u & g_2 Q'_S v_s & M_{\tilde{Z}'} 
 \end{array} \right ) \;\; .
\end{equation}
The basis $(\tilde{B},\,\tilde{W}^3,\, \tilde{h}_d^0,\,\tilde{h}_u^0,\,
\tilde{S},\,\tilde{Z}' )^T$ is rotated into mass eigenstates 
$(\tilde{\chi}^0_1,\tilde{\chi}^0_2,\tilde{\chi}^0_3,\tilde{\chi}^0_4,\tilde{\chi}^0_5,\tilde{\chi}^0_6 )^T$ by 
\begin{equation}
(\tilde{B},\,\tilde{W}^3,\, \tilde{h}_d^0,\,\tilde{h}_u^0,\,
\tilde{S},\,\tilde{Z}' )  = 
(\tilde{\chi}^0_1,\,\tilde{\chi}^0_2,\, \tilde{\chi}^0_3,\,\tilde{\chi}^0_4,\,
\tilde{\chi}^0_5,\,\tilde{\chi}^0_6 )  \; N
\end{equation}
and $N {\cal M}_N N^T ={\rm diag} ( M_{\tilde{\chi}^0_1},M_{\tilde{\chi}^0_2},M_{\tilde{\chi}^0_3},M_{\tilde{\chi}^0_4},
M_{\tilde{\chi}^0_5},M_{\tilde{\chi}^0_6})$ arranged in ascending order. 
$N$ is a 6 by 6 orthogonal matrix since the neutralino mass matrix ${\cal M}_N$ is real and symmetric.
The interactions between the CP-even Higgs boson and a pair of neutralinos
are given by Eq.~(\ref{inte1}) with the corresponding VEV replaced by
$\phi$ (i.e. $v_{u,d,s} \longrightarrow \phi_{u,d,s}$). We can then rotate 
into mass 
eigenstates using (\ref{higgs}) and the interaction terms are given by
\begin{equation}
{\cal L}_{\rm neutralinos}^{\rm int} =\frac{1}{2}\, h_k \, \overline{\tilde{\chi}^0_i}\,
 \left[ {\cal H}^*_{ijk} P_L + {\cal H}_{ijk} P_R \right ]\,\tilde{\chi}^0_j  \;\; ,
\end{equation}
with
\begin{eqnarray}          
{\cal H}_{ijk} &=& 
  {\cal O}_{k1} \left[ \frac{h_s}{\sqrt{2}} N_{i5} N_{j4} 
                     +\frac{e}{2 c_w} N_{i1} N_{j3}
                     -\frac{g}{2} N_{i2} N_{j3}
                     - g_2 Q'_{H_d} N_{i6} N_{j3}  \right ] \nonumber \\
&+& {\cal O}_{k2} \left[ \frac{h_s}{\sqrt{2}} N_{i5} N_{j3} 
                     -\frac{e}{2 c_w} N_{i1} N_{j4}
                     +\frac{g}{2} N_{i2} N_{j4}
                     - g_2 Q'_{H_u} N_{i6} N_{j4}  \right ] \nonumber \\
&+& {\cal O}_{k3} \left[ \frac{h_s}{\sqrt{2}} N_{i4} N_{j3} 
                     - g_2 Q'_{S} N_{i6} N_{j5}  \right ] \nonumber \\
 &+& \{ i \leftrightarrow j \} \; \; .
\end{eqnarray}

\section{Scanning of Parameter Space}

Besides the usual MSSM parameters of gaugino masses $ M_{1,2,3}$, 
squark masses $M_{\tilde{q}}$, slepton masses $M_{\tilde{\ell}}$, $A$
parameters $A_{t,b,\tau}$, and $\tan\beta$, the UMSSM has the following 
additional soft parameters: $M_S$, $M_{\tilde{Z}'}$, $A_s$, 
the VEV $ \langle S \rangle = v_s/\sqrt{2}$, and the Yukawa coupling $h_s$.  
The effective $\mu$ parameter is given as
$\mu_{\rm eff} = h_s \langle S \rangle$.  The other model parameters
are fixed by the quantum numbers $Q'_{\phi}$ of various super-multiplets
$\phi$ as given in Table \ref{table-models}.
The $\eta$ model of $E_6$ defined by the generator in
Eq.~(\ref{eta-model}) or by 
the fifth column for the ${\bf 27}$ in Table \ref{table-models} will be 
used in the following for illustration.

Ignoring the $Z-Z'$ mixing, the mass of the $Z'$ boson is determined by
$m_{Z'} \approx {g_2} ( Q'^2_{H_u} v_u^2 +Q'^2_{H_d} v_d^2 + Q'^2_{s} v_s^2 )^{1/2}$.
The most stringent limit on the $Z'$ boson comes from the dilepton 
resonance search by ATLAS \cite{atlas}. The limits 
are $1.5-1.7$ TeV for the various $Z'$ bosons of the $E_6$ models.
If the limits are translated into $v_s$ using the above expression, the value
of $v_s$ has to be larger than a few TeV. Nevertheless, we can avoid 
these $Z'$ mass limits by assuming the leptonic decay mode is suppressed.
The mixing between the SM $Z$ boson and the $Z'$ can also be suppressed
by carefully choosing the $\tan\beta \approx (Q'_{H_d}/Q'_{H_u})^{1/2}$.
The goal of this work does not concern avoiding all these constraints,
but we note that we can always carefully choose the set of quantum 
numbers $Q'$ such that the $Z'$ mass and mixing constraints can be
evaded.

We first fix most of the MSSM parameters (unless stated otherwise):
\begin{eqnarray}
&& M_{1} = 0.5 M_{2} = 0.2 \;{\rm TeV}, \;\; M_3 = 2 \; {\rm TeV} \; ; \nonumber \\
&& M_{\tilde{Q}} = M_{\tilde{U}} = A_{t} = 1\;{\rm TeV},\;\;
   M_{\tilde{L}} = M_{\tilde{E}} = 0.2 \;{\rm TeV}\; .
\end{eqnarray}
We also fix the following two UMSSM parameters
\begin{equation}
 M_S = 0.5 \,{\rm TeV},\;\; A_s = 0.5 \, {\rm TeV} \;\; ,
\end{equation}
while we scan the rest of the parameters in the following ranges
\begin{equation}
0.2 \,{\rm TeV} <  v_s  < 2 \;{\rm TeV},\;\;
    0.2 < h_s < 0.7 ,\;\;  1.1 < \tan \beta < 40\; , \;
0.2 \, {\rm TeV} < M_{\tilde{Z}'}  <  2 \, {\rm TeV} \; \; .
\label{scan}
\end{equation}

\subsection{Constraints}

{\it Charginos Mass.} The chargino sector of the UMSSM is the same as
that of MSSM with the following chargino mass matrix
\begin{equation}
 M_{\tilde{\chi}^\pm} = \left ( \begin{array}{cc}
        M_2 & \sqrt{2} m_W \sin\beta \\
        \sqrt{2} m_W \cos\beta & \mu_{\rm eff} \end{array} \right ) \;.
\end{equation}
Thus, the two charginos masses depend on $M_2$,
$\mu_{\rm eff}=h_s v_s/\sqrt{2}$, and $\tan\beta$. 
The current bound is $M_{\tilde{\chi}^\pm}> 94$ GeV as long as
the mass difference with the lightest supersymmetric particle (LSP) 
is larger than 3 GeV \cite{pdg}. 
We impose this chargino mass bound in our scans in the parameter space
defined by (\ref{scan}).

{\it Invisible Width of the $Z$ Boson.} The lightest neutralino 
$\tilde{\chi}^0_1$ is the LSP of the model, and thus would be stable
and invisible.  When the $Z$ boson decays into a pair of LSP, it 
would  give rise to invisible width of the $Z$ boson, which had been
tightly constrained by experiments. The current bound of the $Z$ invisible
width is $\Gamma_{\rm inv} (Z) < 3$ MeV at about 95\% CL \cite{pdg}.
The coupling of the $Z$ boson to the lightest neutralino is given by
\begin{equation}
{\cal L}_{Z\tilde{\chi}_1^0 \tilde{\chi}^0_1} = 
  \frac{g_1}{4} \, \left( | N_{13}|^2 - | N_{14} |^2 \right ) \,
Z_\mu \, \overline{\tilde{\chi}^0_1} \gamma^\mu \gamma_5
 \, \tilde{\chi}^0_1\; ,
\end{equation}
and the contribution to the $Z$ boson invisible width is
\begin{equation}
\Gamma(Z \to \tilde{\chi}^0_1 \tilde{\chi}^0_1 ) =
  \frac{g_1^2} { 96 \pi} \left( | N_{13}|^2 - | N_{14} |^2 \right )^2  m_Z
 \left( 1 - \frac{4 m_{\tilde{\chi}^0_1}^2 } { m_Z^2 } \right )^{3/2} \; .
\end{equation}
Here we impose the experimental constraint on the invisible $Z$ width.
The constraint of fulfilling the relic density by the LSP will be ignored
in this work.

Current limits on the pseudoscalar Higgs bosons come from the LEP searches
of $e^+ e^- \to Z^* \to A_i H_j$, where $i,j$ denote the mass eigenstates 
of the Higgs bosons; especially in those MSSM-extended models, such as
NMSSM, with multiple pseudoscalar and scalar Higgs bosons the constraint could
be severe.  However, there is only one pseudoscalar Higgs boson in the UMSSM
and in our choice of parameters it is often heavier than a few hundred GeV.
Thus, it is not constrained by the current limits.  Similarly, the charged
Higgs boson is also heavy and not constrained by current searches.

\subsection{The First Scenario: $ 130 < M_{h_{\rm SM-like}}  < 141$ GeV}

In Summer 2011, the LHC experiments reported a $2\sigma$ excess in
the channel $h \to W W^* \to \ell^+ \nu \ell^- \bar \nu$ ($\ell
=e,\mu$) above the expected SM backgrounds, the implied Higgs boson
mass is around $130-141$ GeV and the branching ratio into $WW^*$ is about
$1/2$ of the SM value \cite{recent}.
Nevertheless, in December 2011 the most updated data \cite{126-atlas,126-cms} indicated
that SM Higgs boson above 131 GeV up to about 600 GeV is ruled out.  
It does not mean that a Higgs boson in the mass range above 131 GeV cannot
exist, but just we have to find some ways to hide the Higgs boson. 
Therefore, when we scan for the SM-like Higgs boson in the mass range
$130-141$ GeV, we also search for the region that allows this Higgs boson
to be invisible. We shall elaborate further about this below.

\begin{figure}[th!]
\centering
\includegraphics[angle=270,width=4in]{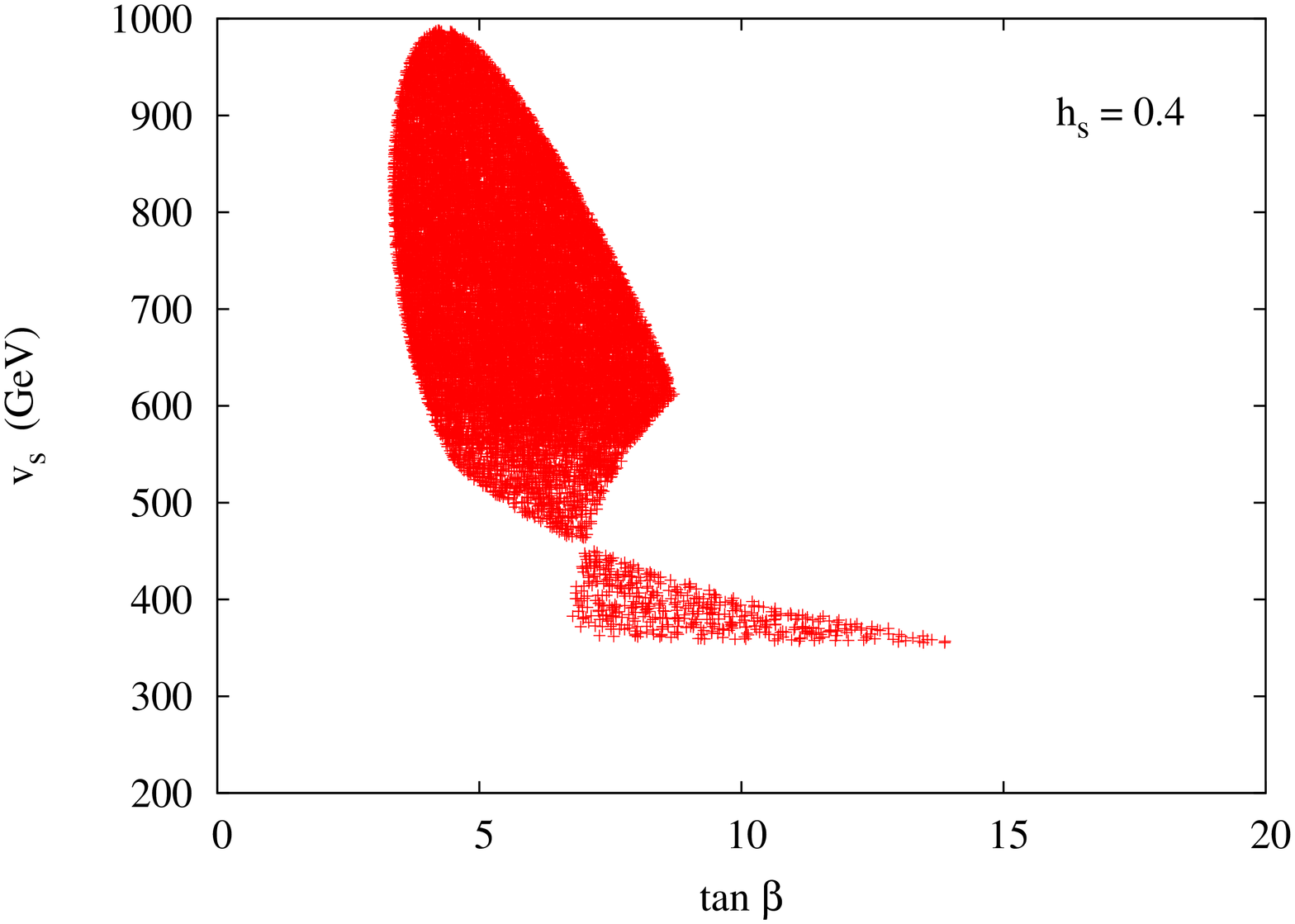}
\includegraphics[angle=270,width=4in]{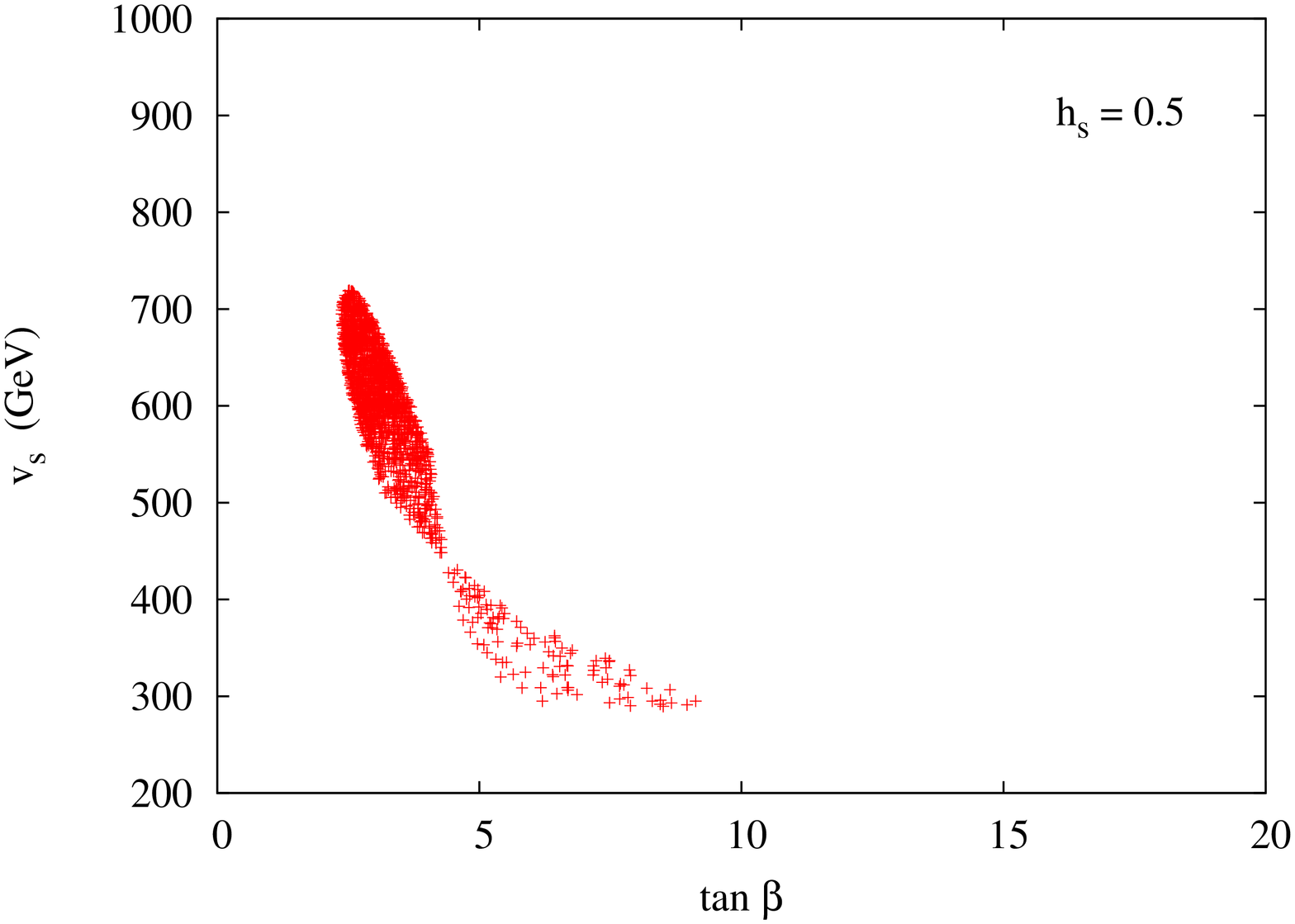}
\includegraphics[angle=270,width=4in]{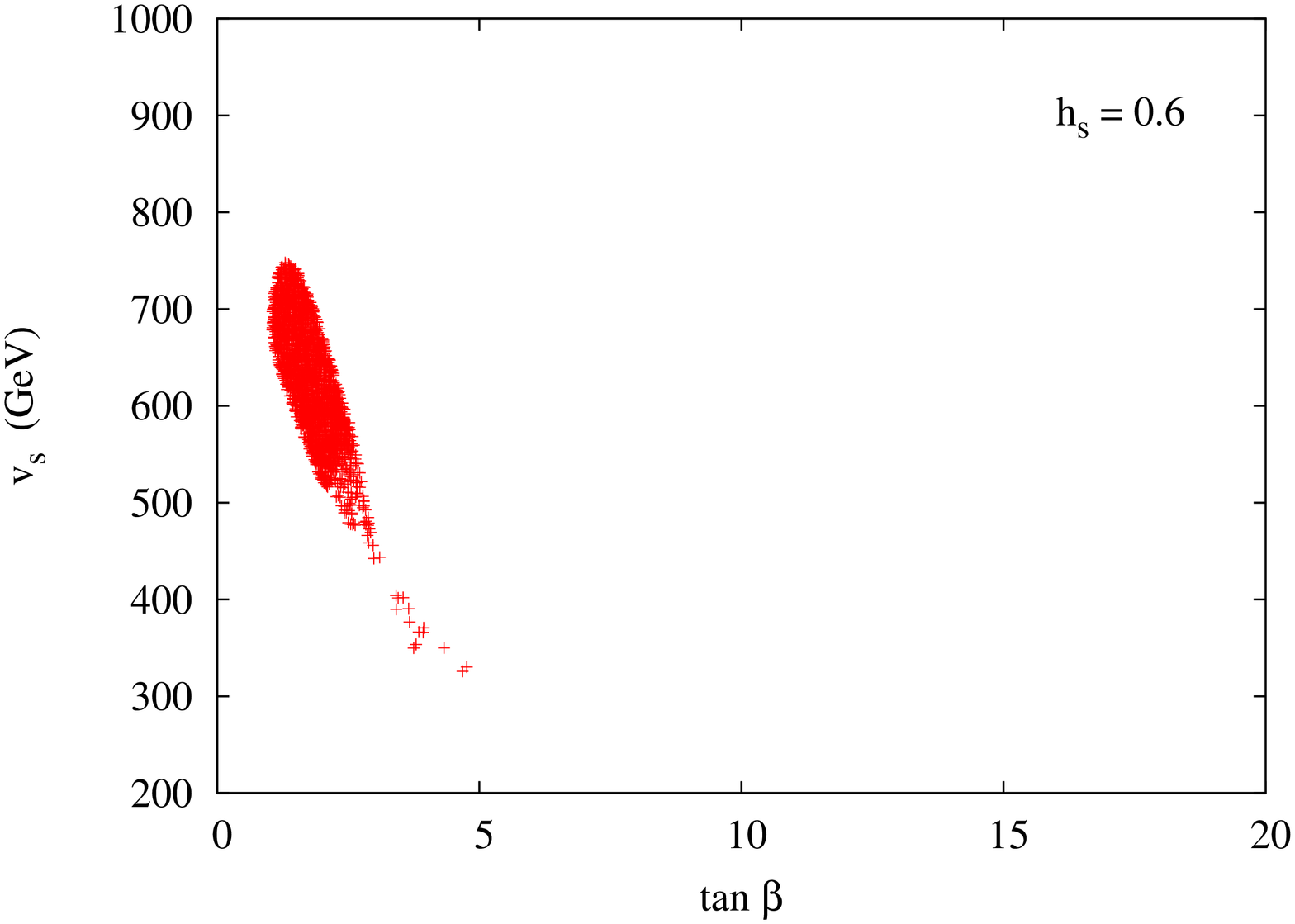}
\caption{\small \label{fig1}
Two-dimensional scatter plots for the parameter-space points satisfying 
the chargino mass constraint $M_{\tilde{\chi}^\pm} > 94$ GeV, 
invisible $Z$ width less than 3 MeV, and
$ 130 < M_{h_{\rm SM-like}} < 141$ GeV, where 
the SM-like Higgs boson $h_{\rm SM-like}$ satisfies $O_{k3}^2 < 0.1$ 
(where $h_k = O_{k1} \phi_d + O_{k2} \phi_u + O_{k3} \phi_s$).
}
\end{figure}

We first do the parameter space scan to
search for the points that can give a SM-like Higgs boson of mass
between 130 and 141 GeV.  Here the SM-like Higgs boson is {\it not}
always the lightest CP-even Higgs boson. Sometimes, the lightest Higgs
boson is the singlet-like Higgs boson.  We define the Higgs boson
$h_k$ to be SM-like by demanding the $O_{k3}^2 < 0.1$ (where $h_k =
O_{k1} \phi_d + O_{k2} \phi_u + O_{k3} \phi_s$). 
In our scan, we do not find more than one SM-like Higgs bosons.
We show the points that pass the constraints of chargino mass,
invisible $Z$ width, and the mass of the SM-like Higgs boson 
between 130 and 141 GeV in Figs.~\ref{fig1} for
a number of $h_s$ values.  It is obvious from the figure that 
a smaller $h_s$ is more likely to give a Higgs boson in the mass
range $130-141$ GeV.  The $v_s$ is between 300 GeV and 1 TeV, and 
$\tan\beta$ is between 2 and 15.  The variation of $M_{\tilde{Z}'}$ 
in our selected range is rather uniform and thus no preferred range
of $M_{\tilde{Z}'}$.

\begin{figure}[t!]
\centering
\includegraphics[angle=270,width=2in]{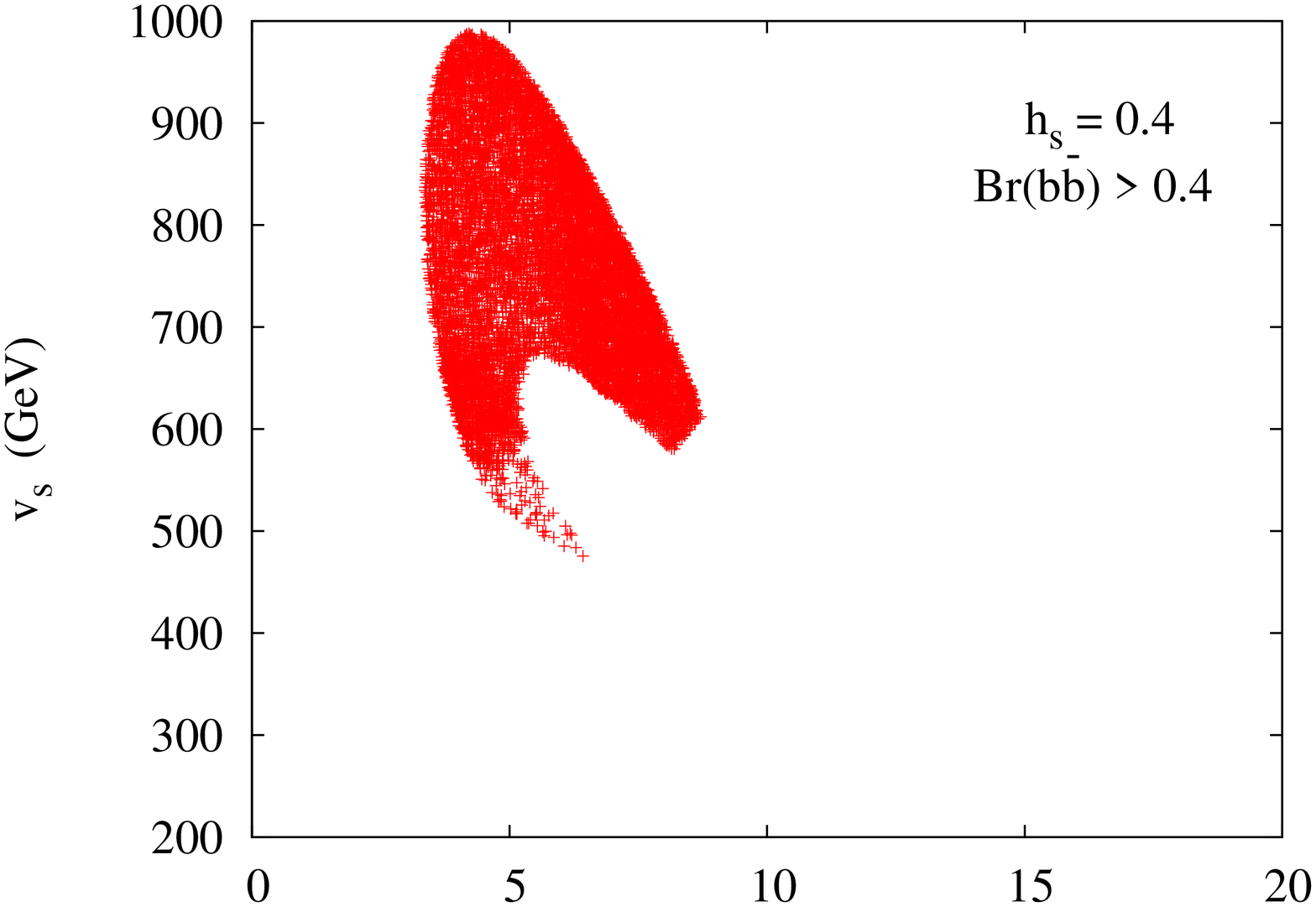}
\includegraphics[angle=270,width=2in]{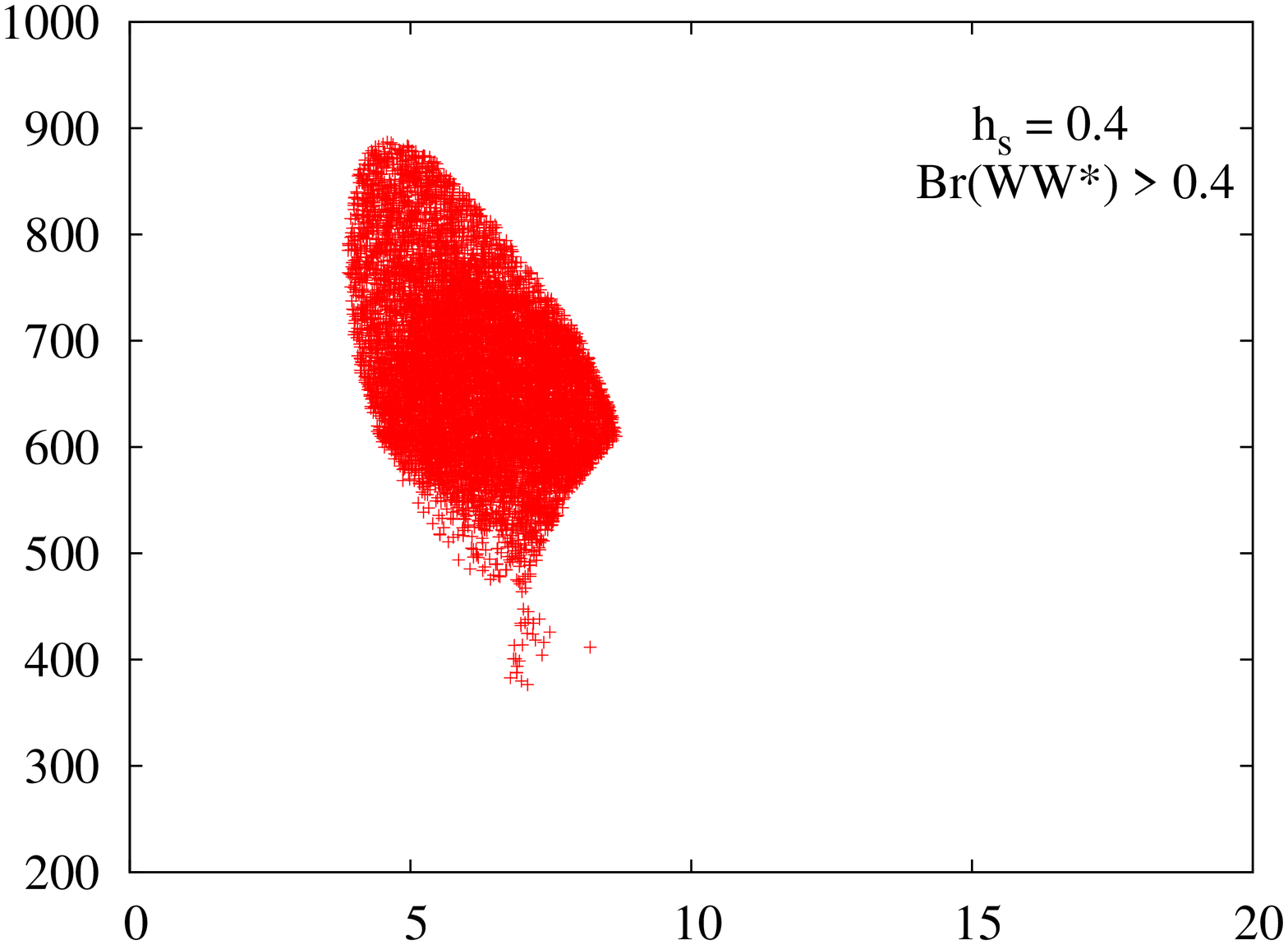}
\includegraphics[angle=270,width=2in]{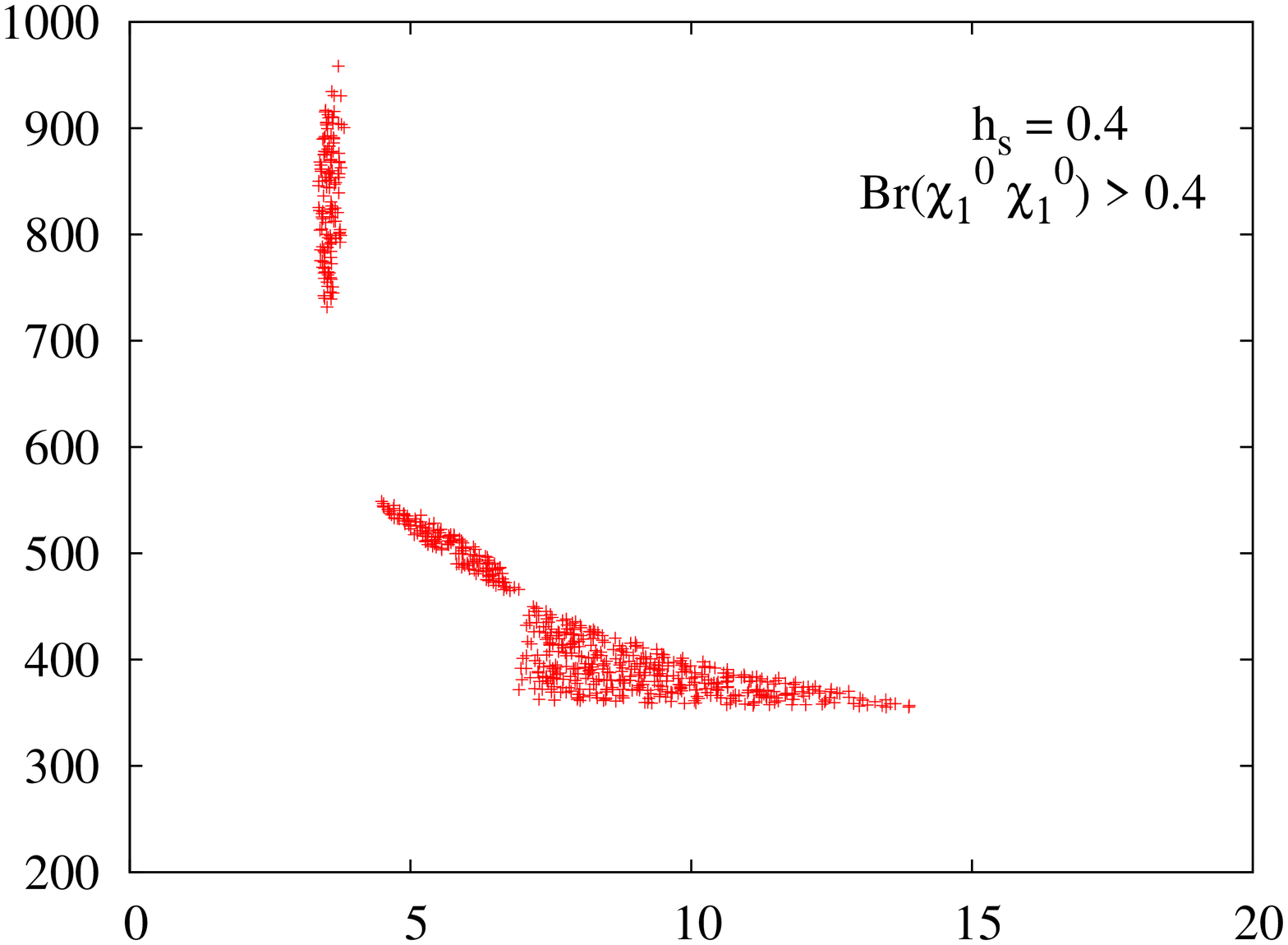}
\includegraphics[angle=270,width=2in]{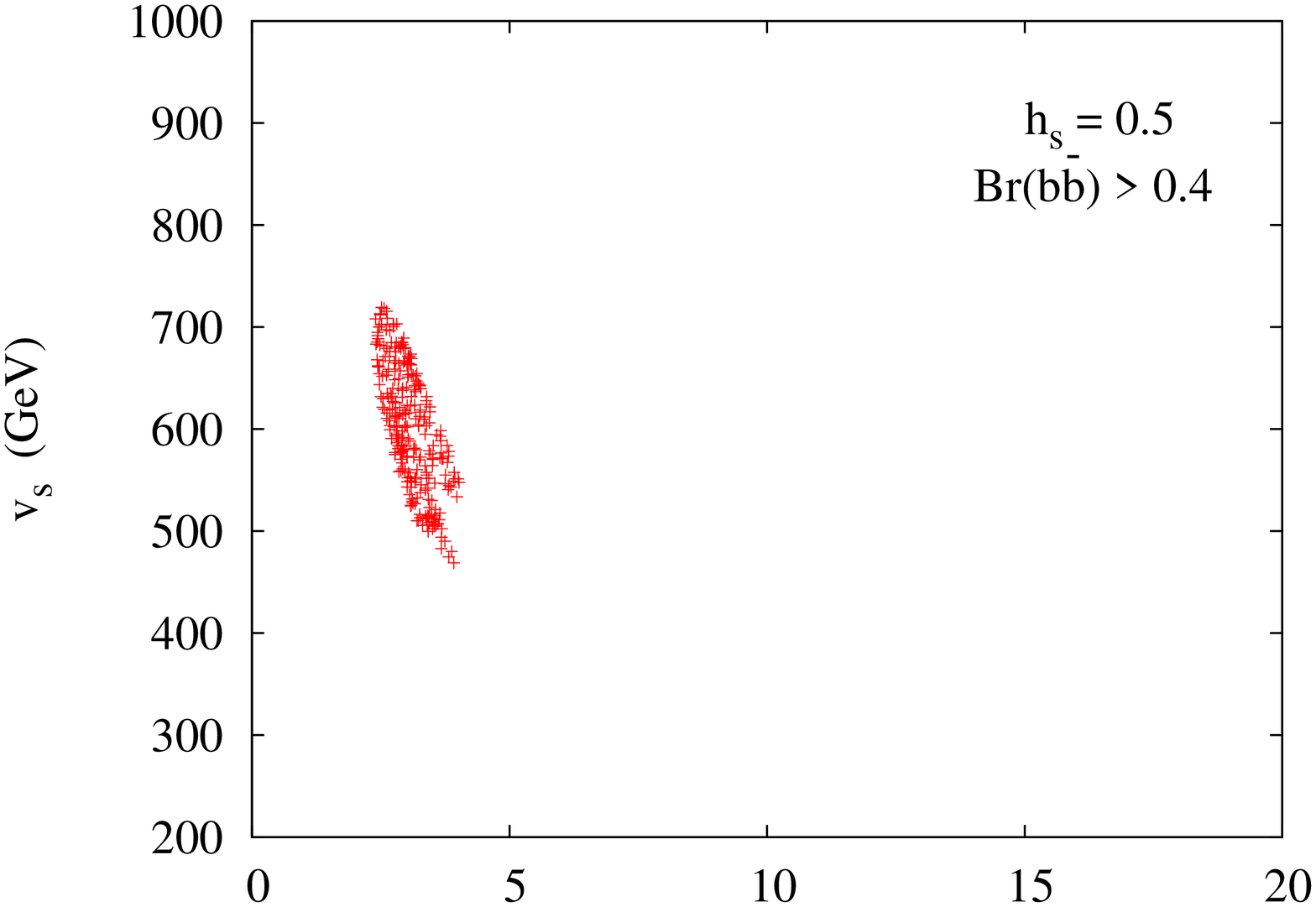}
\includegraphics[angle=270,width=2in]{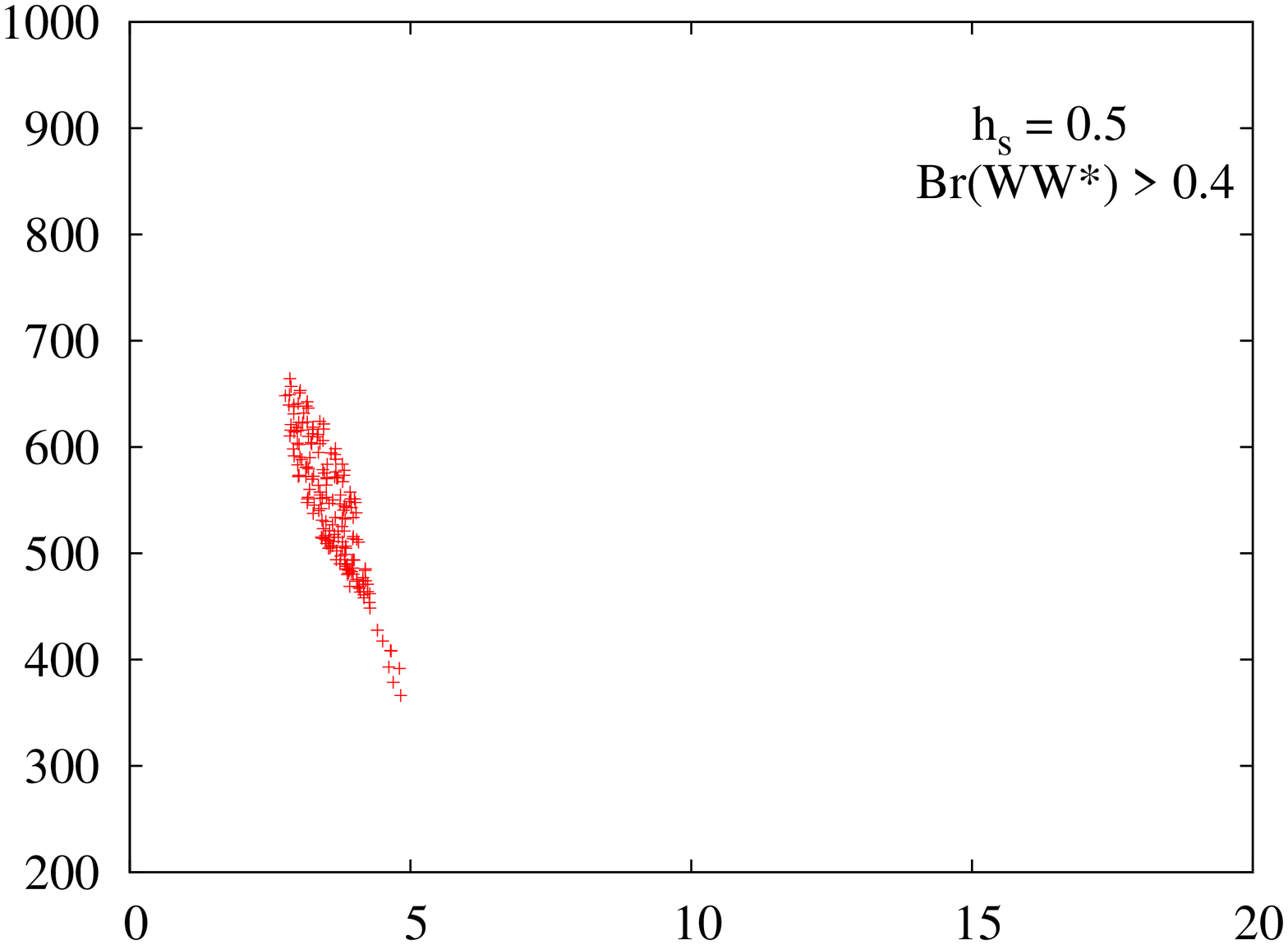}
\includegraphics[angle=270,width=2in]{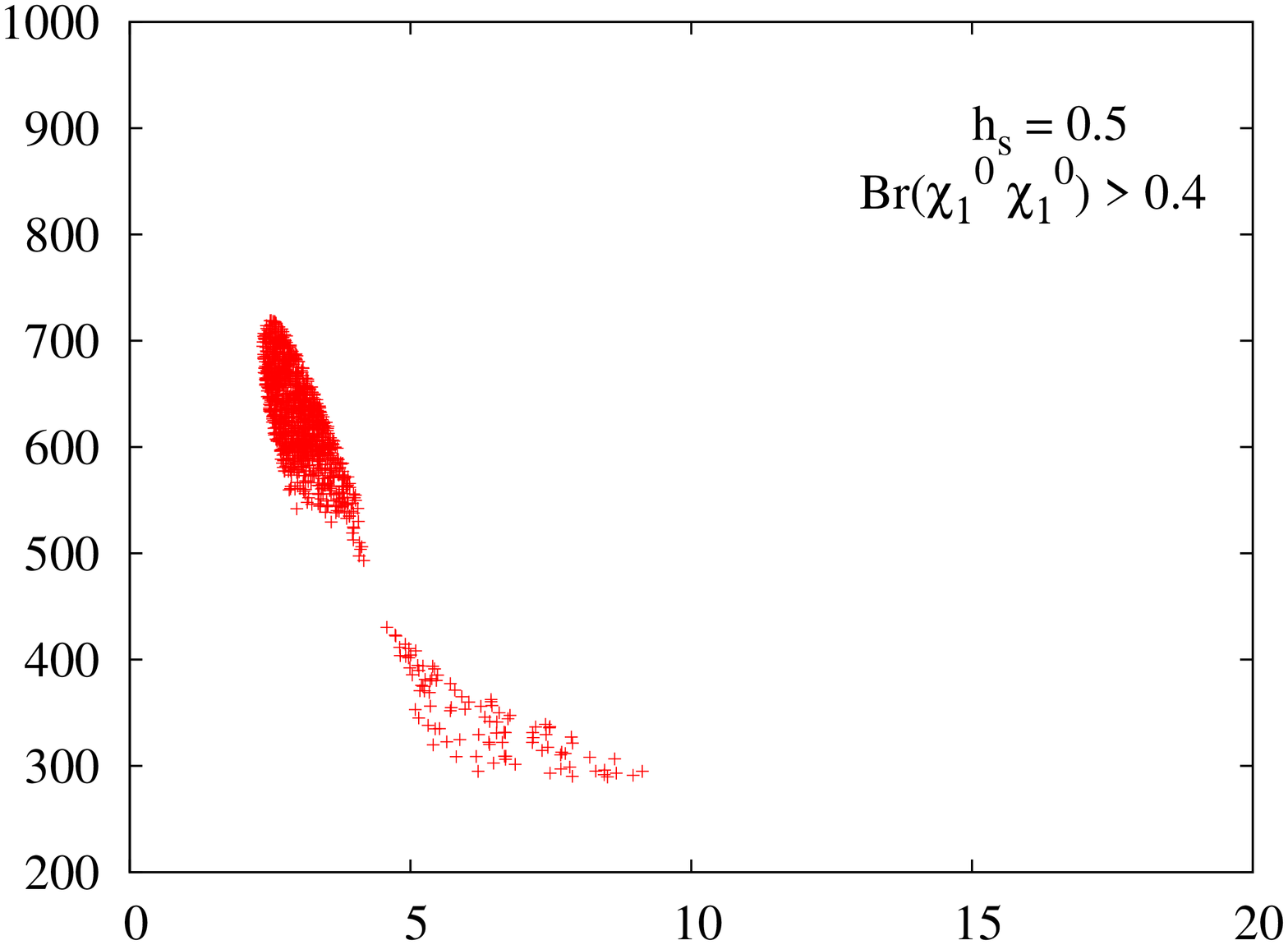}
\includegraphics[angle=270,width=2in]{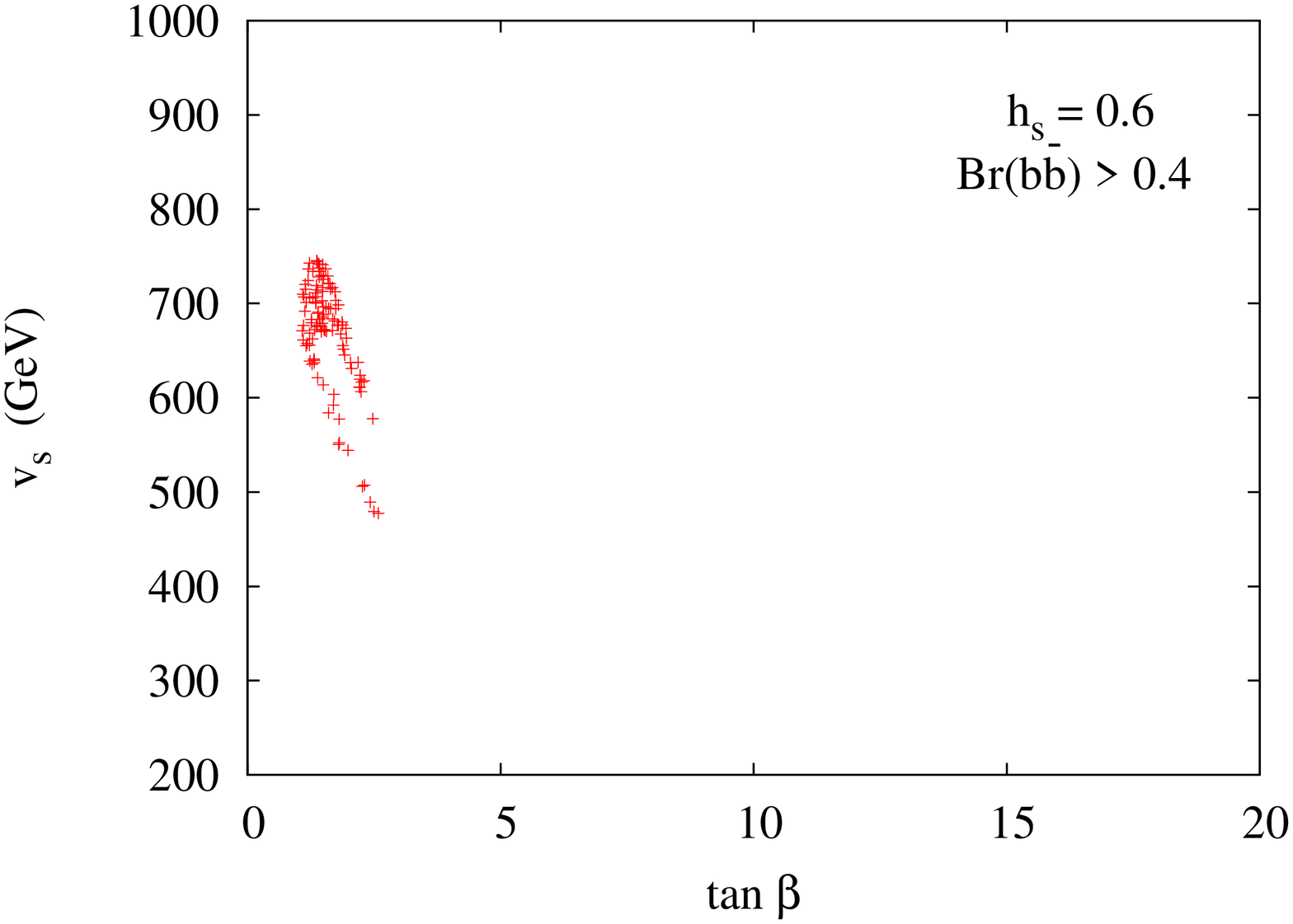}
\includegraphics[angle=270,width=2in]{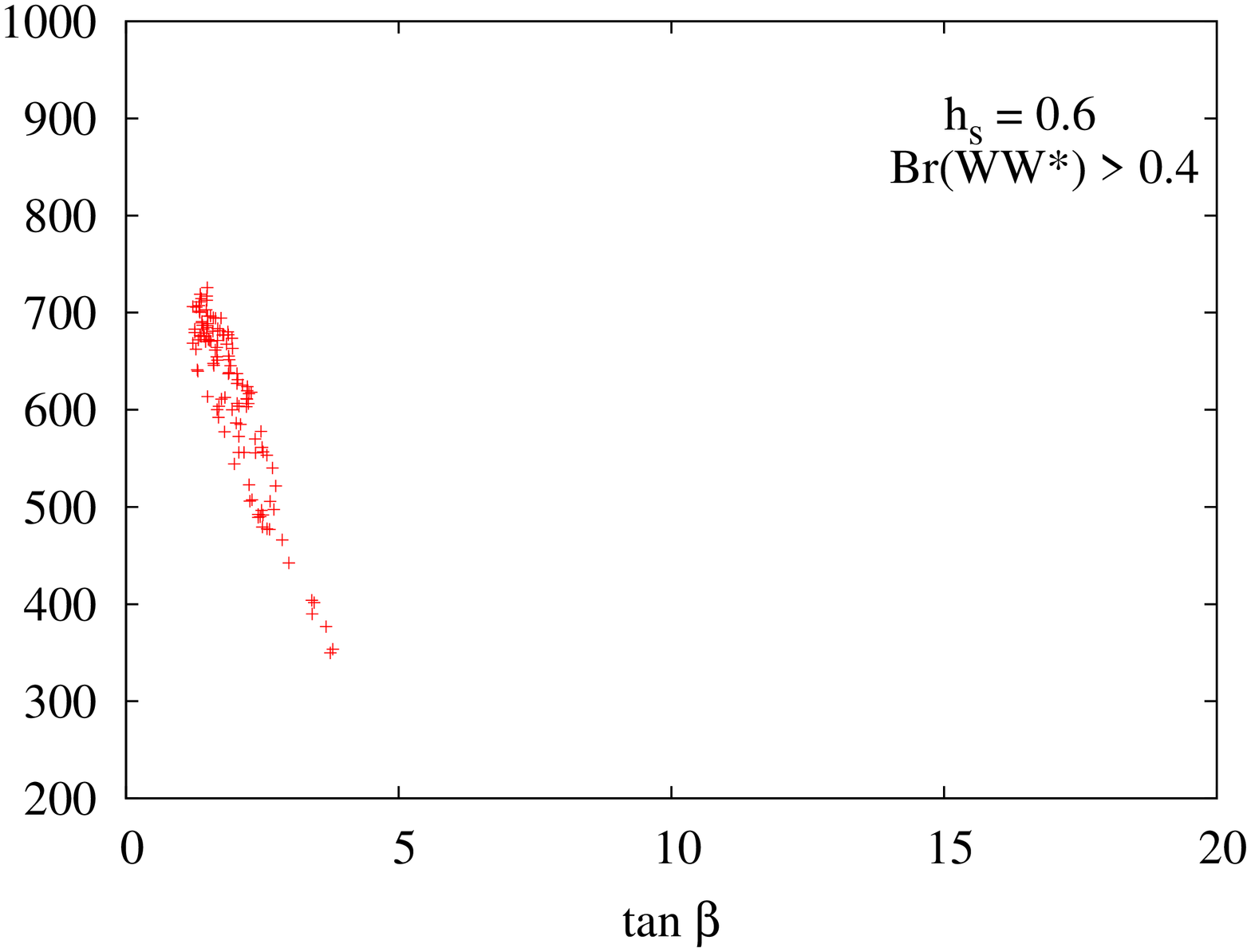}
\includegraphics[angle=270,width=2in]{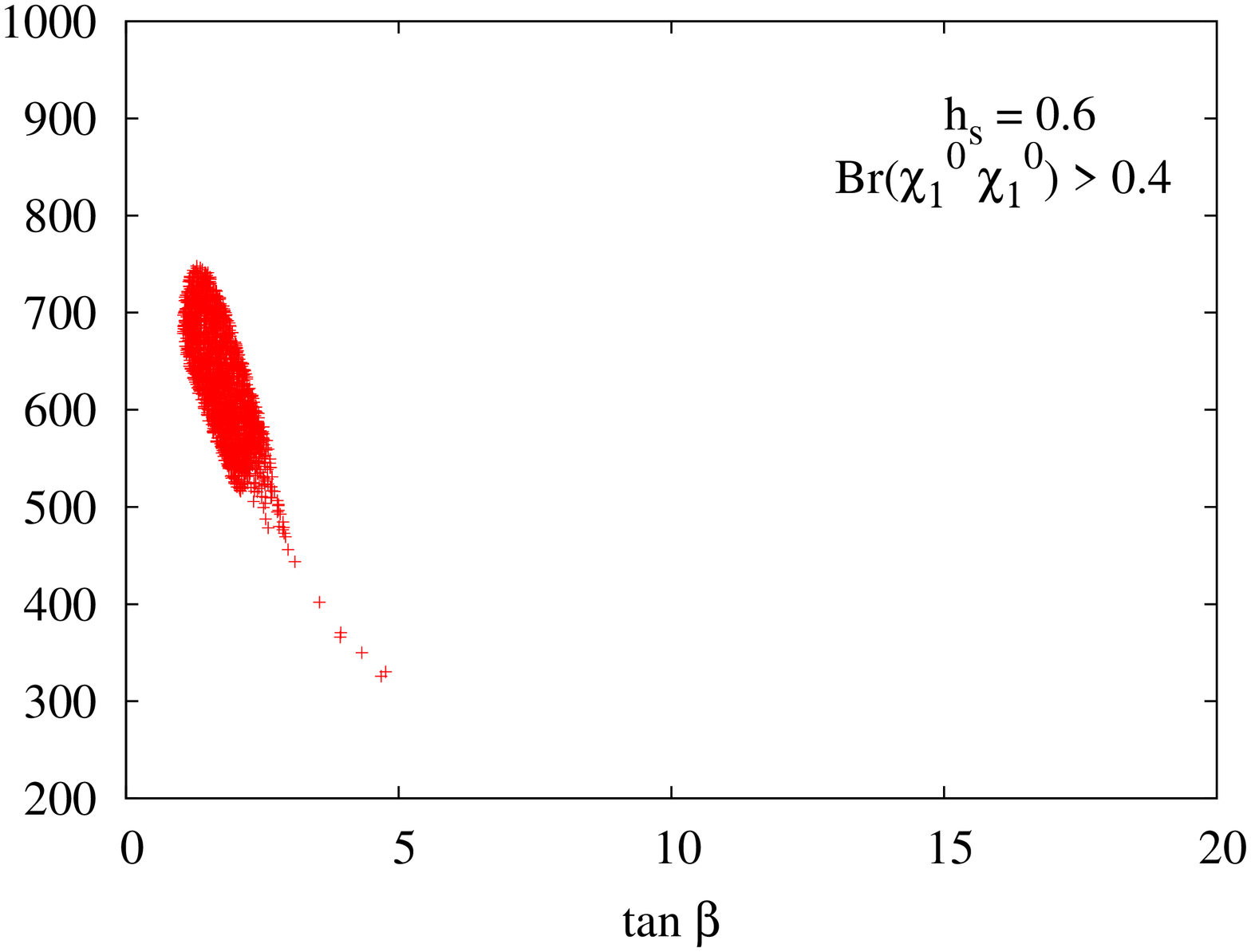}
\caption{\small \label{fig2}
These scattered plots first pass the requirements of charginos masses 
$M_{{\tilde \chi}^\pm} >$ 94 GeV,
the invisible $Z$ width $\Gamma_{\rm inv}(Z) <$ 3 MeV, and 
$130\, {\rm GeV} < M_{h_{\rm SM-like}} < 141\;
{\rm GeV}$.
The first row for $h_s =0.4$, the second row for $h_s=0.5$, and
the third row for $h_s=0.6$. The first column for 
$B( h_{\rm SM-like} \to b\bar b) > 0.4$,
the second column for $B( h_{\rm SM-like} \to WW^*) > 0.4$, and
the third column for $B( h_{\rm SM-like} \to 
\tilde{\chi}^0_1 \tilde{\chi}^0_1) > 0.4$.
}
\end{figure}

Once we have obtained the points with $M_{h_{\rm SM-like}}$ between
130 and 141 GeV, we can then calculate the branching ratios. In the mass
range $130-141$ GeV, the dominant decay modes of the SM-like Higgs boson
include $b\bar b$, $\tau^+ \tau^-$, $W W^*$, $ZZ^*$, and $\tilde{\chi}^0_1
\tilde{\chi}^0_1$.  Among these decay modes either $b\bar b$, $W W^*$, or 
$\tilde{\chi}^0_1 \tilde{\chi}^0_1$ usually dominate.
We found that if the SM-like Higgs boson is $h_2$, 
$h_2$ is always lighter 
than twice the lightest $h_1$ mass so that $h_2 \to h_1 h_1$
is absent in our scan.  We show in Fig.~\ref{fig2} the parameter space
points obtained in Fig.~\ref{fig1} that have the branching ratio
$B(h_{\rm SM-like} \to b\bar b) > 0.4$ in the first column,
$B(h_{\rm SM-like} \to WW^*) > 0.4$ in the second column, and
$B(h_{\rm SM-like} \to \tilde{\chi}^0_1 \tilde{\chi}^0_1) > 0.4$ in the 
third column.  The rows from top to bottom are for $h_s$ = 0.4, 0.5, and 0.6, respectively.
For smaller $h_s$ the invisible mode is not as frequent as the other 
visible modes ($b\bar b$ and $WW^*$), while for larger $h_s$ the invisible
mode is more frequent. In most recent results of ATLAS \cite{126-atlas} and CMS \cite{126-cms},
the SM Higgs boson above 130 GeV and up to about 600 GeV is ruled out.  
Possible ways out include adding invisible or dijet decay modes to the
Higgs boson.  Therefore, if a Higgs boson has an invisible decay mode
with a branching ratio larger than about $0.4$, it can survive the
search limit from the LHC.  The parameter space points with 
$B(h_{\rm SM-like} \to \tilde{\chi}^0_1 \tilde{\chi}^0_1) > 0.4$ presented 
here can then survive the LHC limits.
So, the current LHC data prefers a larger $h_s$ if the SM-like Higgs boson
falls in the mass range of $130-141$ GeV.

\subsection{The Second  Scenario: $ 120 < M_{h_{\rm SM-like}}  < 130$ GeV}

\begin{figure}[th!]
\centering
\includegraphics[angle=270,width=4in]{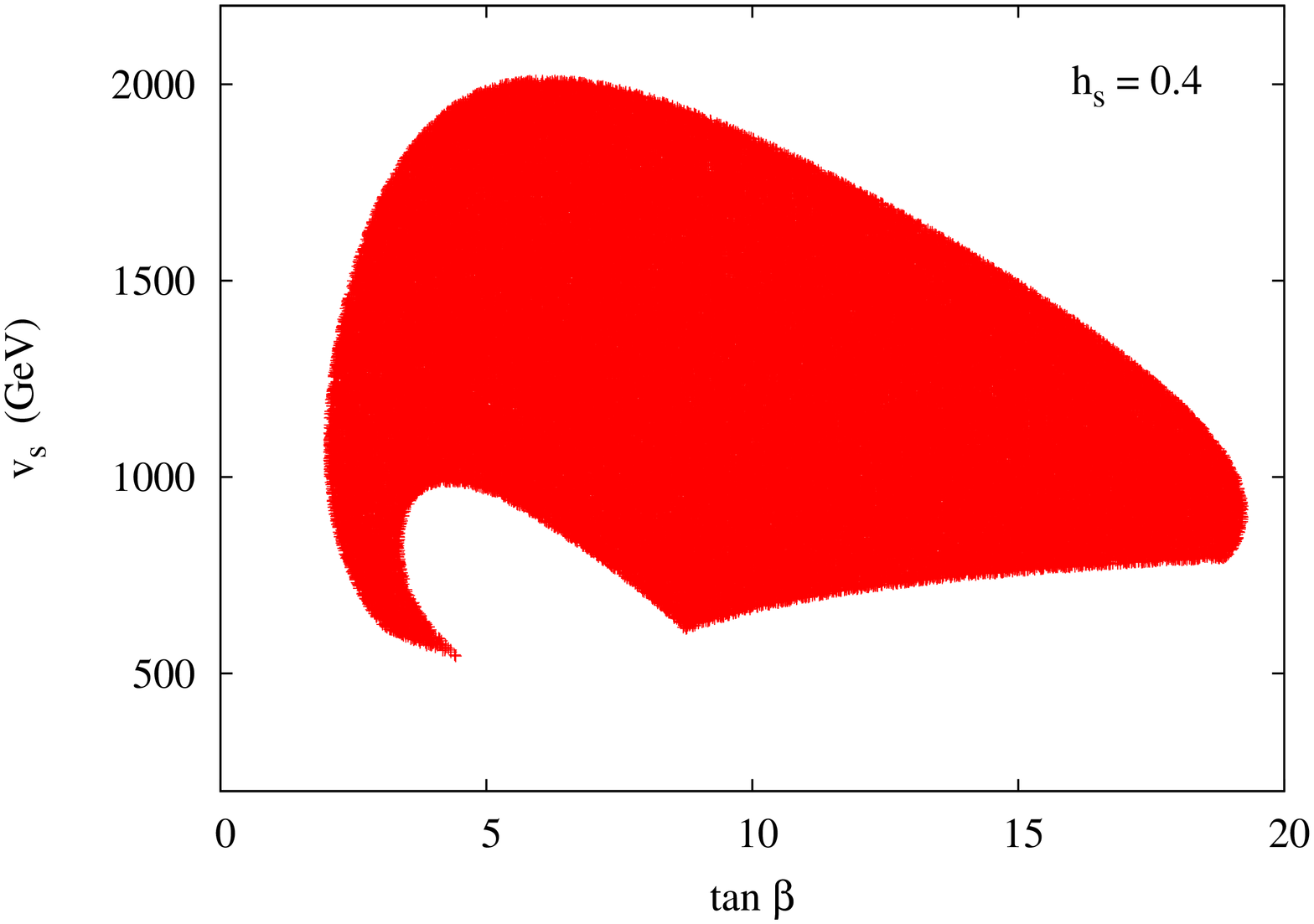}
\includegraphics[angle=270,width=4in]{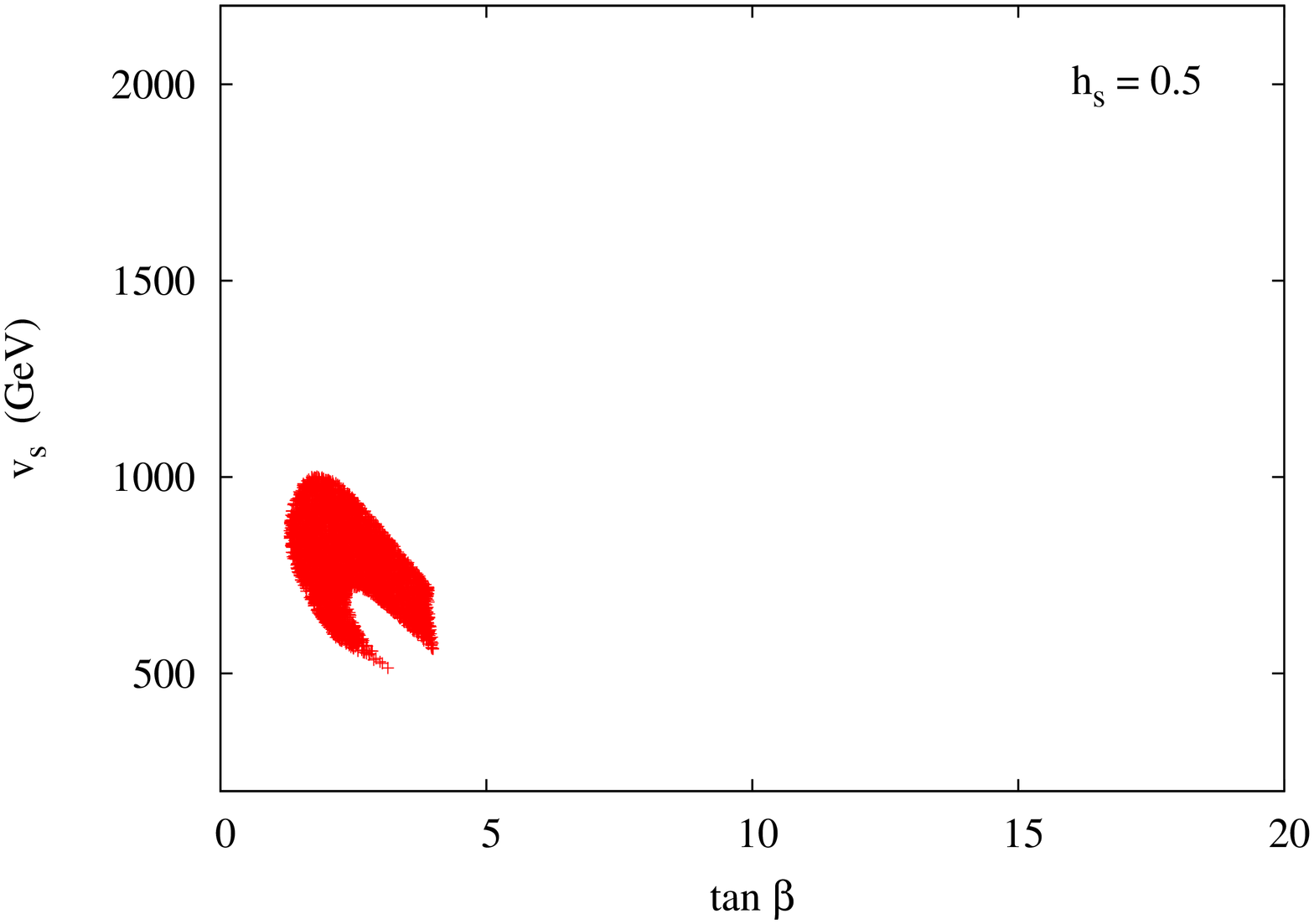}
\includegraphics[angle=270,width=4in]{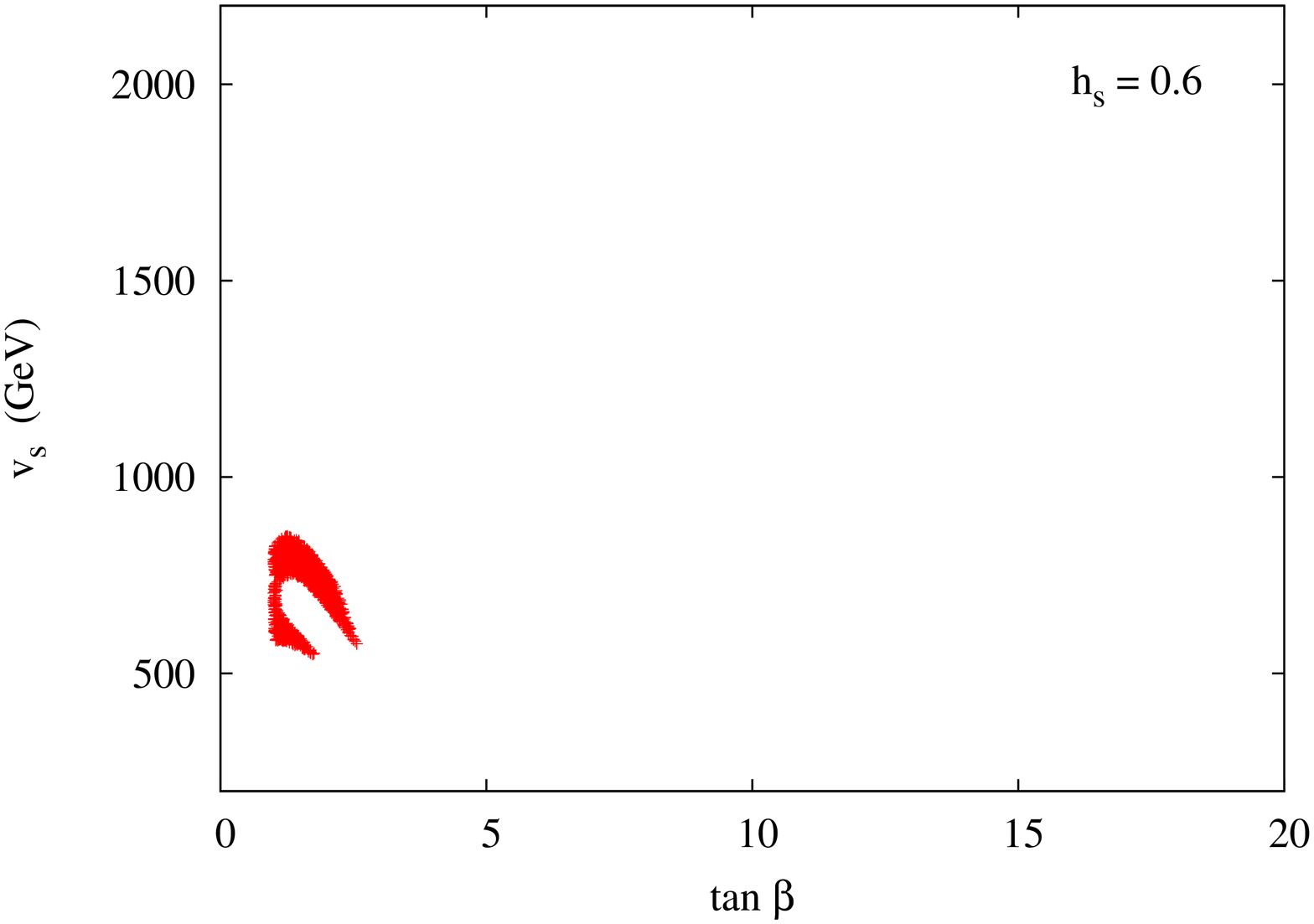}
\caption{\small \label{fig3}
Two-dimensional scatter plots for the parameter-space points satisfying 
the chargino mass constraint $M_{\tilde{\chi}^\pm} > 94$ GeV, 
invisible $Z$ width less than 3 MeV, and
$ 120 < M_{h_{\rm SM-like}} < 130$ GeV, where 
the SM-like Higgs boson $h_{\rm SM-like}$ satisfies $O_{k3}^2 < 0.1$ 
(where $h_k = O_{k1} \phi_d + O_{k2} \phi_u + O_{k3} \phi_s$).}
\end{figure}
\begin{figure}[th!]
\centering
\includegraphics[angle=270,width=2in]{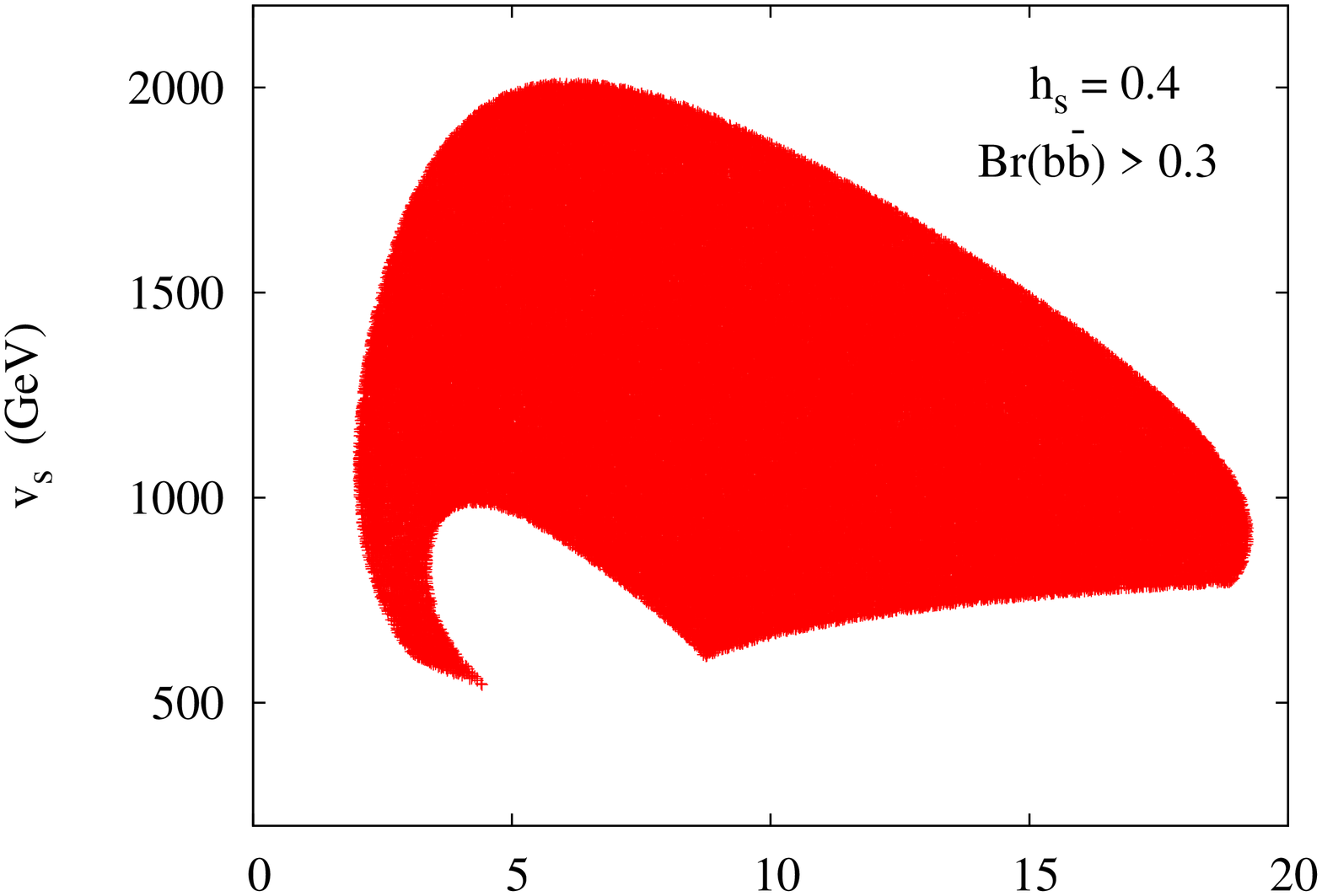}
\includegraphics[angle=270,width=2in]{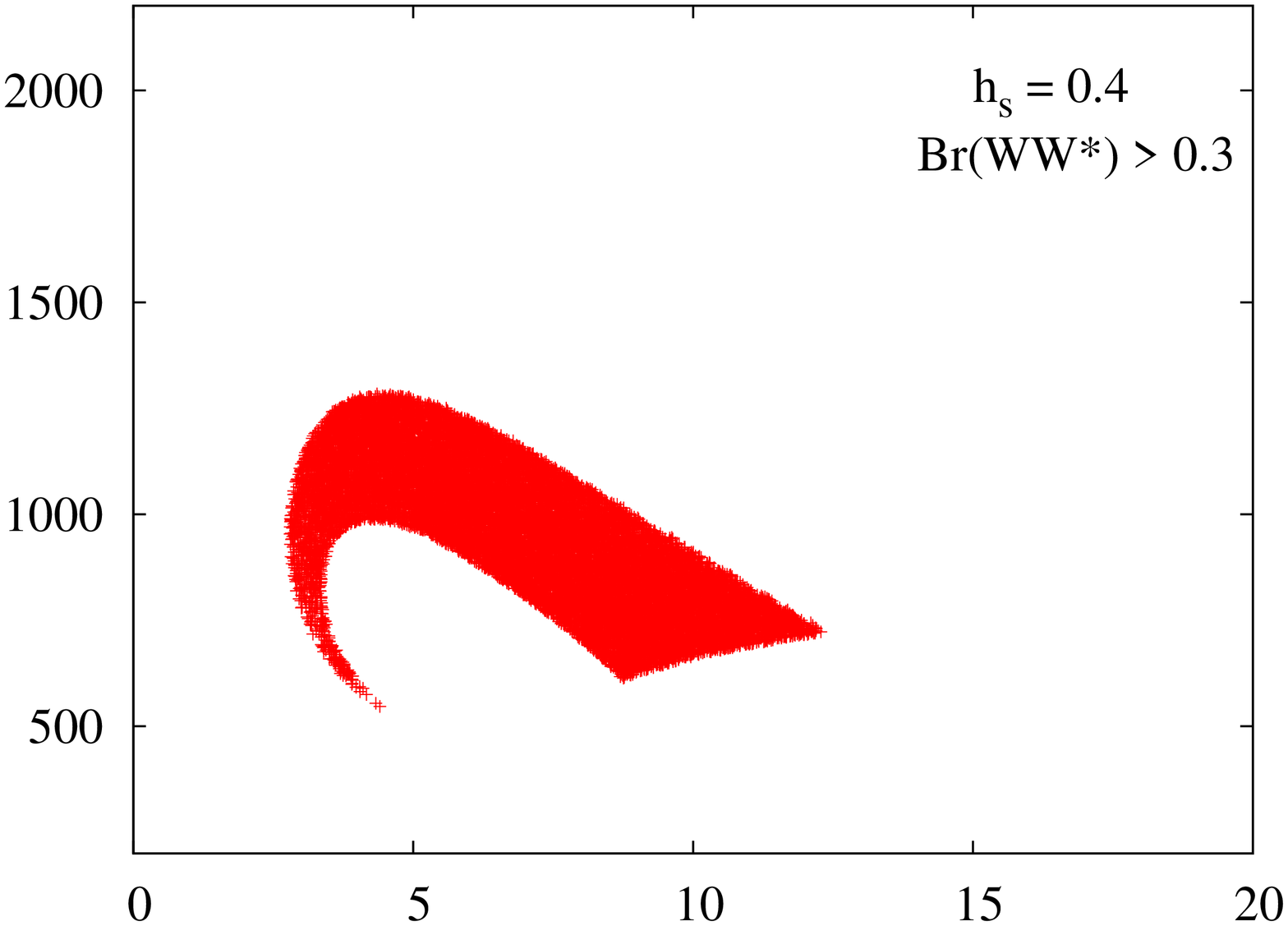}
\includegraphics[angle=270,width=2in]{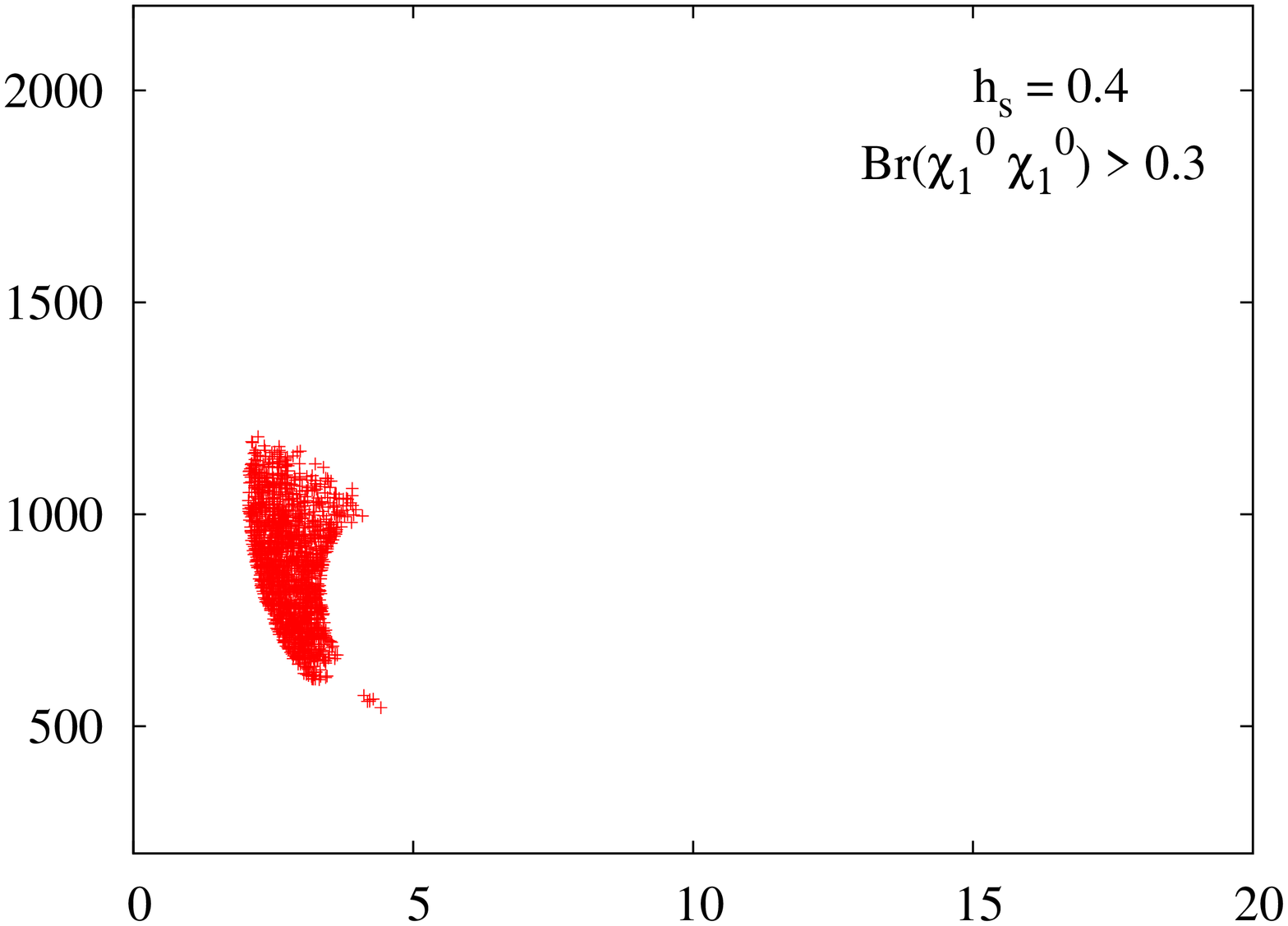}
\includegraphics[angle=270,width=2in]{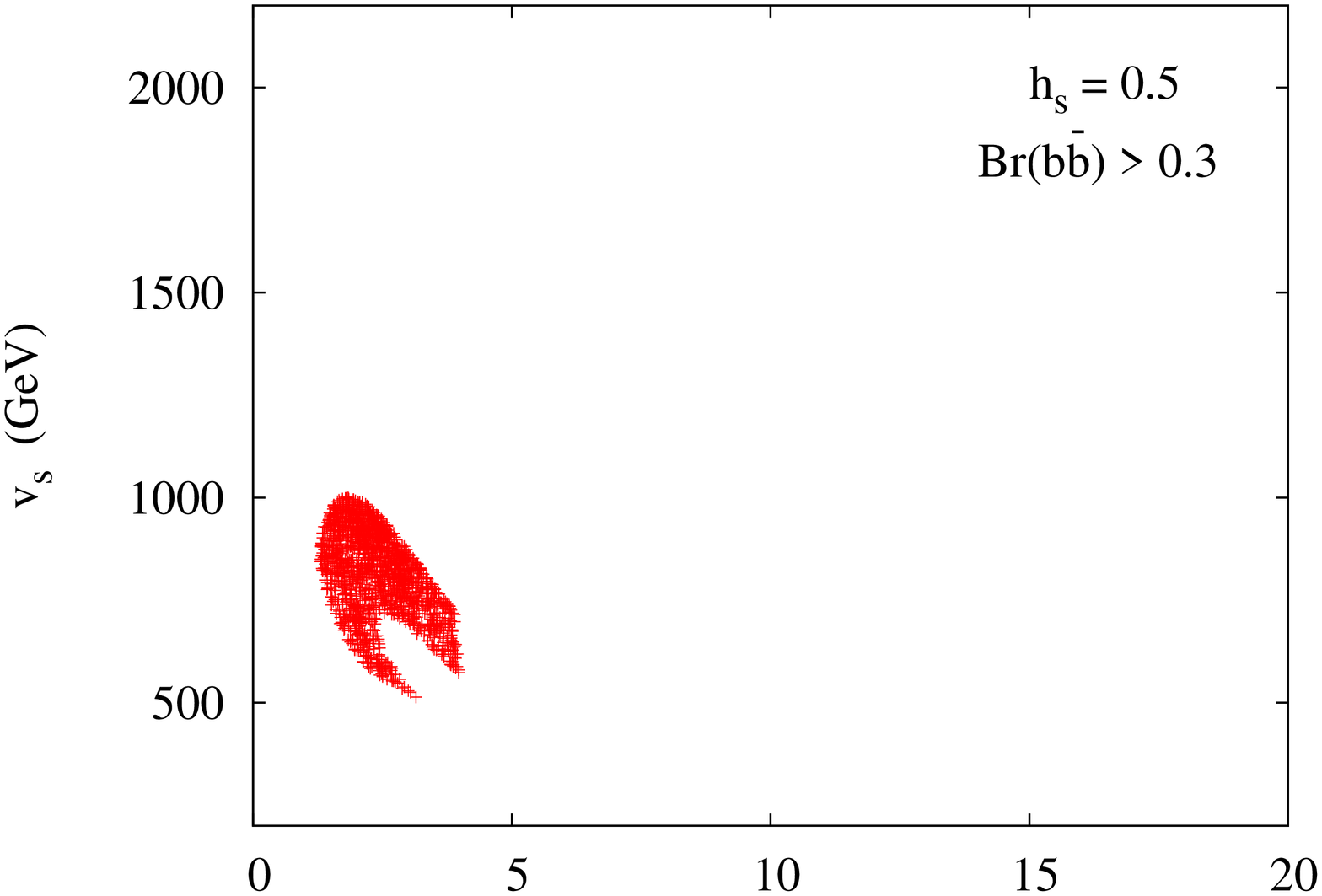}
\includegraphics[angle=270,width=2in]{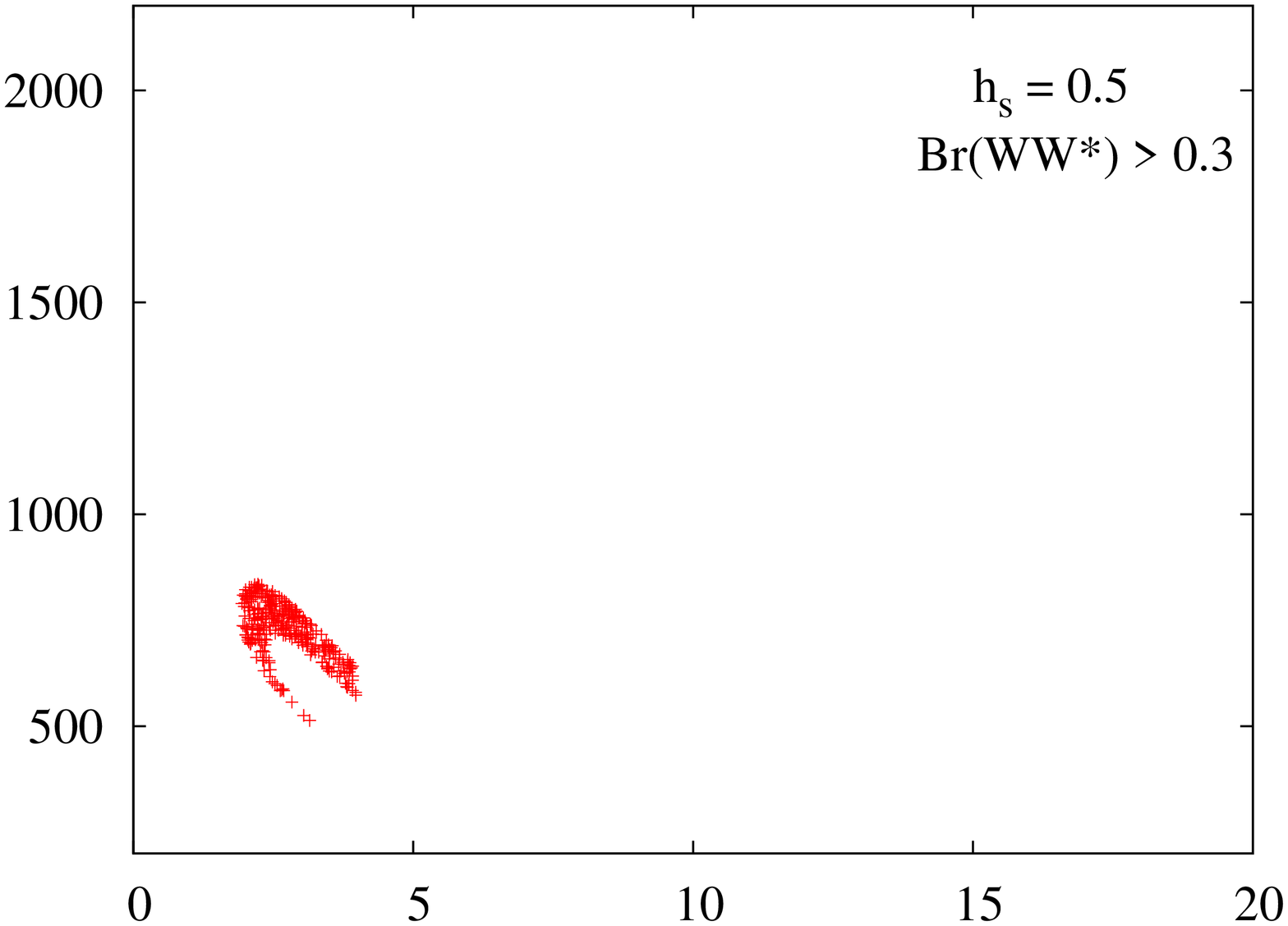}
\includegraphics[angle=270,width=2in]{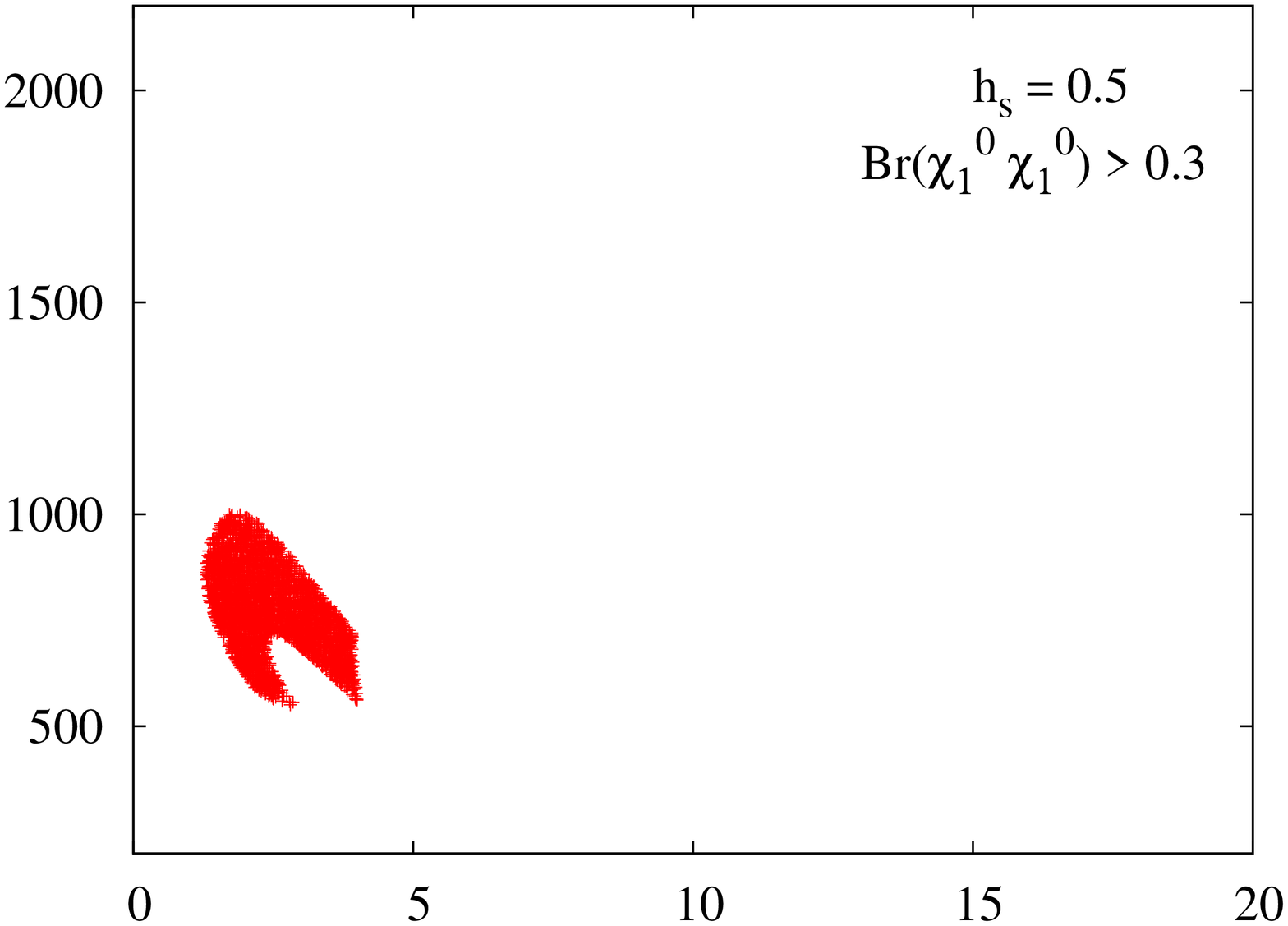}
\includegraphics[angle=270,width=2in]{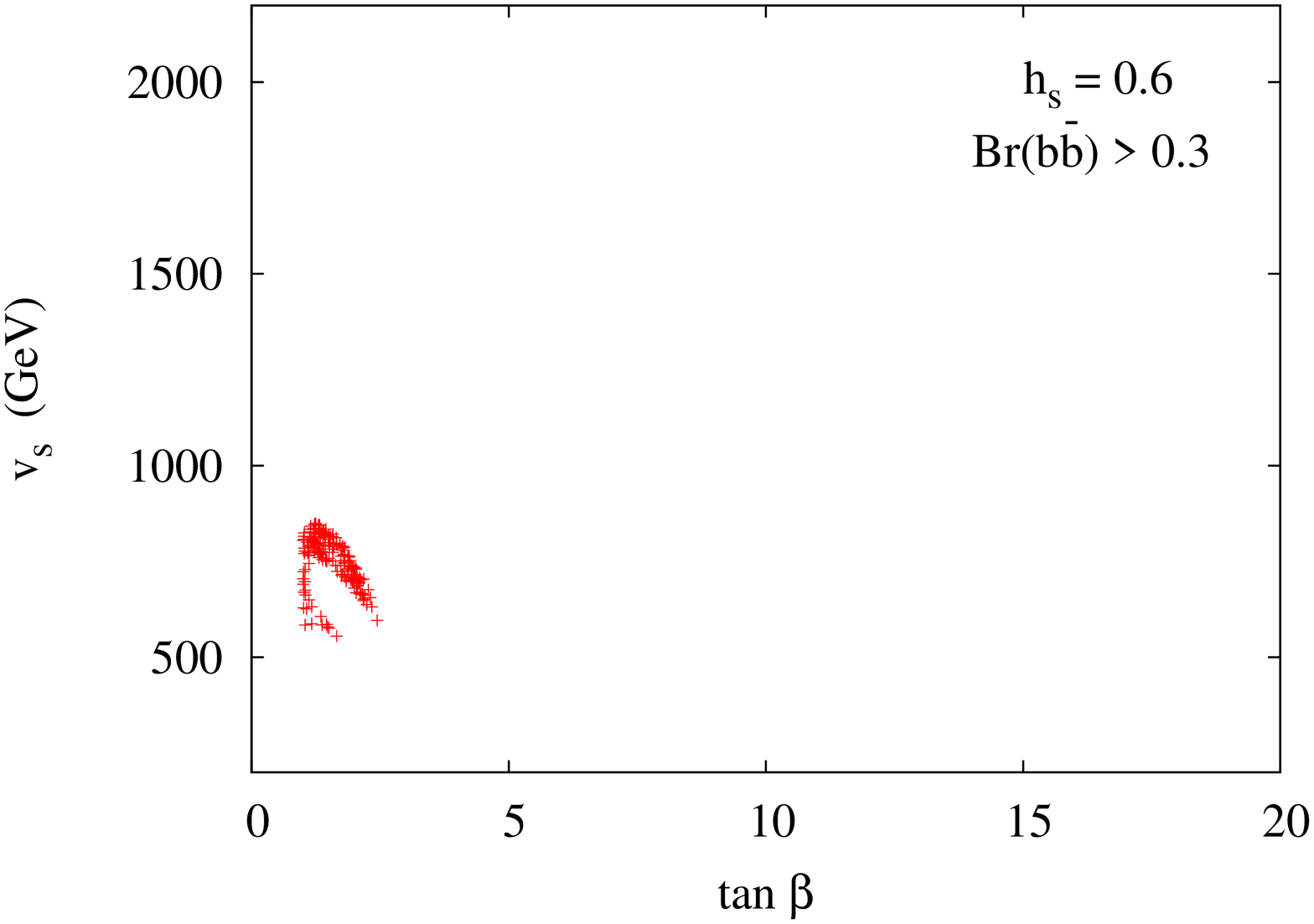}
\includegraphics[angle=270,width=2in]{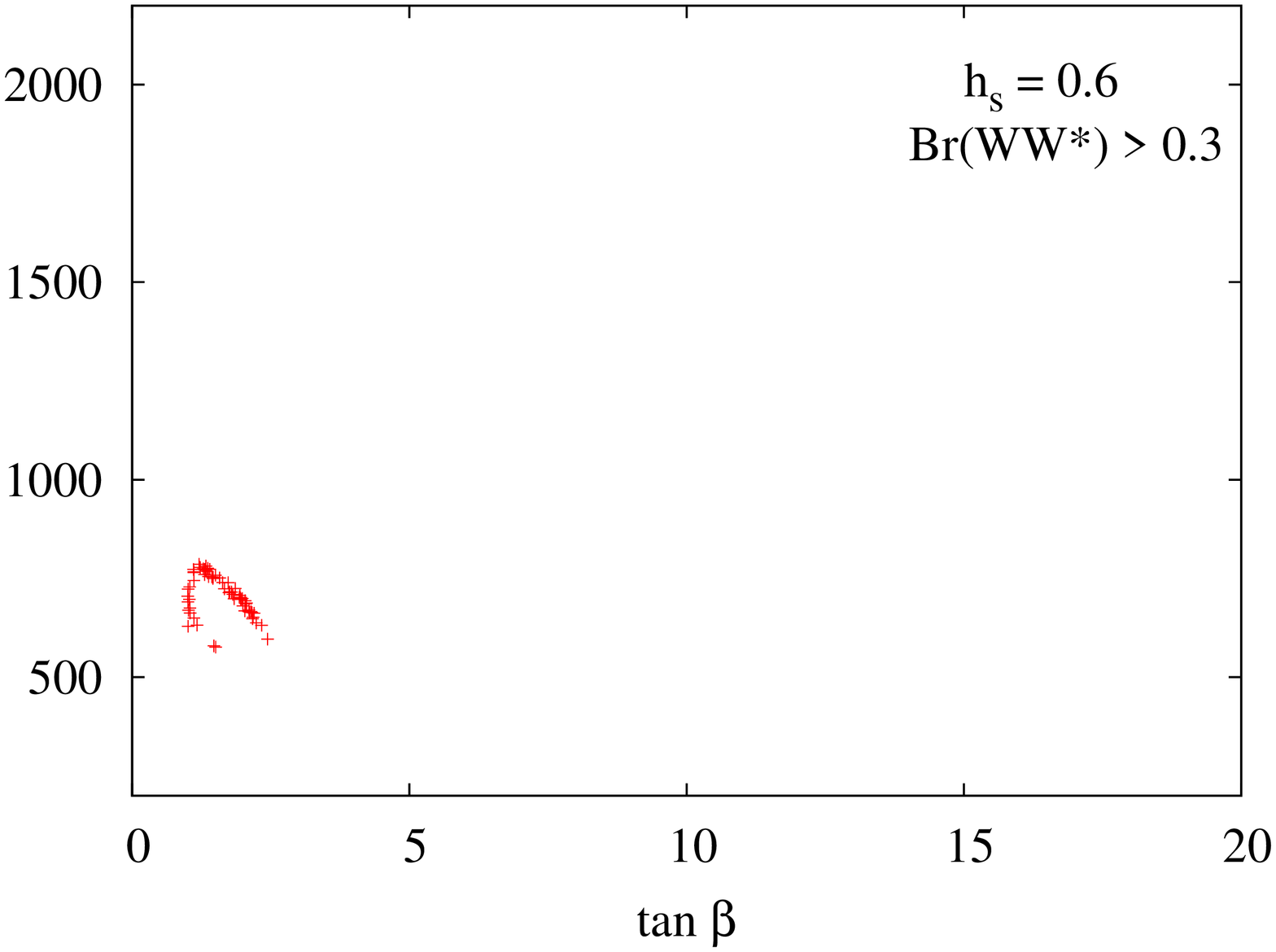}
\includegraphics[angle=270,width=2in]{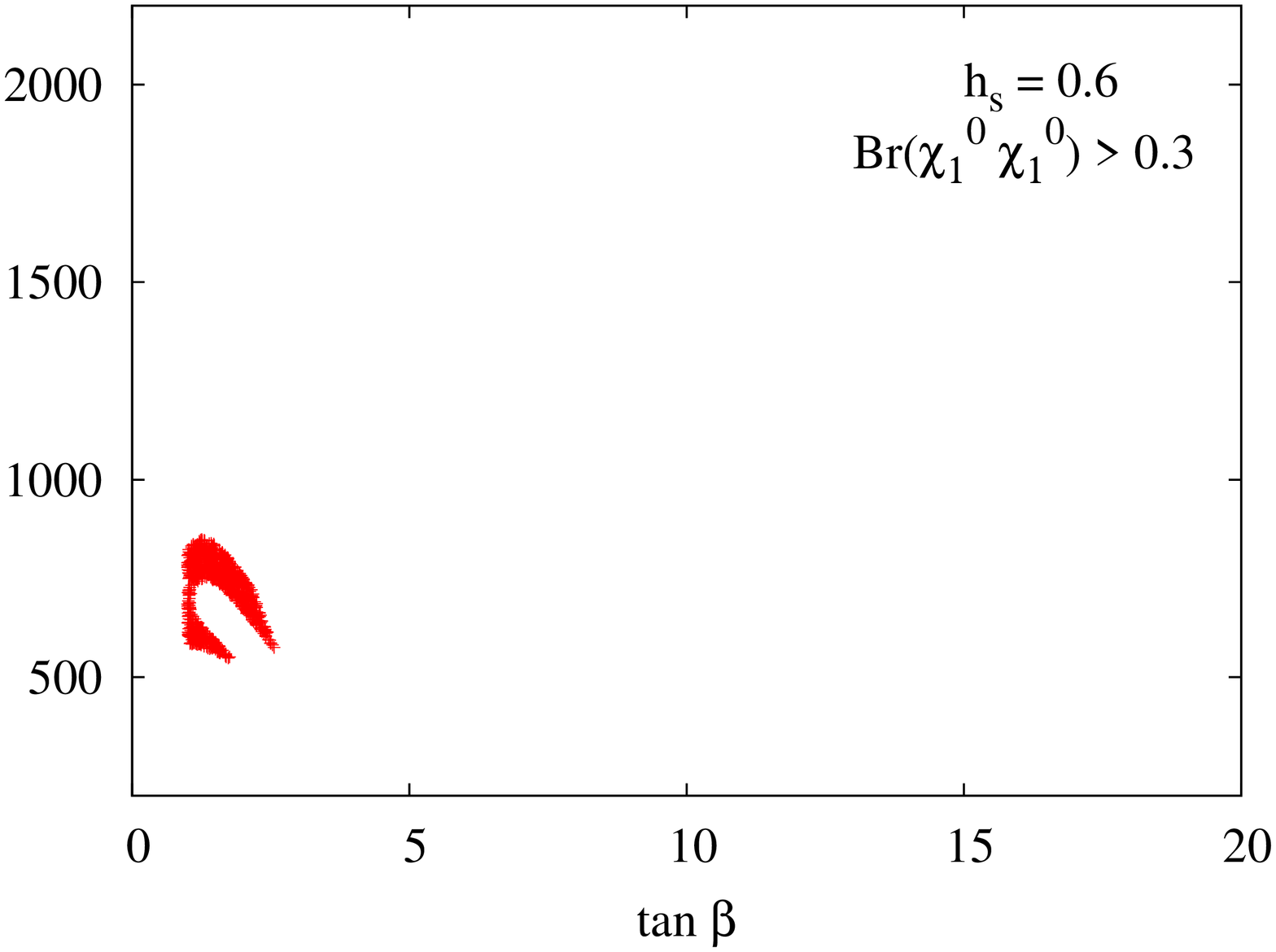}
\caption{\small \label{fig4}
These scattered plots first pass the requirements of charginos masses
$M_{{\tilde \chi}^\pm} >$ 94 GeV,
the invisible $Z$ width $\Gamma_{\rm inv}(Z) <$ 3 MeV, and 
$120\, {\rm GeV} < M_{h_{\rm SM-like}} < 130\;
{\rm GeV}$.
The first row for $h_s =0.4$, the second row for $h_s=0.5$, and
the third row for $h_s=0.6$. The first column for 
$B( h_{\rm SM-like} \to b\bar b) > 0.3$,
the second column for $B( h_{\rm SM-like} \to WW^*) > 0.3$, and
the third column for $B( h_{\rm SM-like} \to 
\tilde{\chi}^0_1 \tilde{\chi}^0_1) > 0.3$.}
\end{figure}

We repeat the whole exercise 
in the previous scenario with the new
requirement of Higgs boson mass
in the range $120 < M_{h_{\rm SM-like}}  < 130$ GeV.  We show the 
parameter space points that satisfy the chargino mass bound, $Z$ invisible
width, and $120 < M_{h_{\rm SM-like}}  < 130$ GeV in Fig.~\ref{fig3}.
It is also true for this mass range that a smaller $h_s$ is easier to give
a SM-like Higgs boson mass of $120-130$ GeV.  For $h_s=0.4$ the $v_s$
extends from 500 GeV to 2 TeV, and $\tan\beta$ from 2 to 18.
For $h_s=0.5,0.6$ the ranges of $v_s$ and $\tan\beta$ are substantially smaller.
In Fig.~\ref{fig4}, we show the parameter-space points that each branching
ratio $B(b\bar b)> 0.3$, $B(WW^*)>0.3$, and 
$B(\tilde{\chi}^0_1 \tilde{\chi}^0_1) > 0.3$.  We used $0.3$ in this figure
because the points for $WW$ and $\tilde{\chi}^0_1 \tilde{\chi}^0_1$ would be
very few if we chose $0.4$.  
At such a low-mass range the $b\bar b$ often dominates over the
$WW^*$, and the $b\bar b$ mode also dominates over 
$\tilde{\chi}^0_1 \tilde{\chi}^0_1$ for $h_s=0.4$;  while for 
$h_s$ = 0.5 and 0.6, the $\tilde{\chi}^0_1 \tilde{\chi}^0_1$ mode is indeed
dominant.  This feature is similar to the other mass range $130-141$ GeV:
when $h_s$ is large the invisible mode becomes more important. 
Therefore, the current LHC data prefers a smaller $h_s$ if the SM-like
Higgs boson falls in the mass range of $120-130$ GeV.

Another feature of the current LHC data showed that the production rate of 
the Higgs boson into diphotons is slightly larger than that of the 
SM Higgs boson \cite{126-atlas,126-cms}. However, one has to be careful 
that the 
current data consists of large statistical uncertainties, and the data are
consistent either with the presence of the
SM Higgs boson or without any Higgs boson. 
It has been shown in a number of recent works that in MSSM 
\cite{mssm} or NMSSM \cite{jack,ell-higgs} the production rate of diphotons is 
similar to that of the SM Higgs boson, mostly slightly smaller than
the SM one, though at some points in the parameter space
it could be slightly larger. 
Nevertheless, under some less restrictive conditions
the production rate of diphotons may be enhanced by up to a factor of 2
in the NMSSM \cite{ell-higgs}.
Here, we do not expect the UMSSM can give a dramatic
change in diphoton production rate, as long as the SM-like Higgs boson
does not decay into the lightest neutralinos. We show those points
that have substantial branching ratios into $b\bar b$ and $WW$
in Fig.~\ref{fig4} (first and second column).  
In this case, the production rate into
diphotons would not be any different from the MSSM predictions,
because the gluon-fusion is very
similar and so is the decay into diphotons, except for a slight singlet
component in the Higgs boson couplings.  
Therefore, in this subsection we have shown the
parameter space of UMSSM that can give a SM-like Higgs boson
of mass  $120-130$ GeV with branching ratios similar to those of
the SM Higgs boson.  On the other hand, we also show the 
parameter space points that the SM-like Higgs boson decays
mostly into invisible neutralinos in the last column of Fig.~\ref{fig4}.

\section{Discussion}

In principle, in both scenarios studied in the previous section, 
there may be some parameter space that the second lightest Higgs 
boson is SM-like and can also decay into the lightest Higgs boson, which
is mostly singlet-like. However, in our scan we do not find such 
parameter-space points.   

{\it Comparison with the SM Higgs boson.} If the SM Higgs boson falls
in the mass range of larger than 130 GeV, it would be inconsistent 
with the current data \cite{126-atlas,126-cms}. The UMSSM, on the other hand, can
allow the SM-like Higgs boson in this mass range to decay invisibly 
into the lightest neutralinos, such that it can hide from the 
current data. The SM-like Higgs boson in the UMSSM
can also accommodate in the lighter mass range of
$120-130$ GeV with the decay branching ratios very similar to those in the SM.

{\it Comparison with the MSSM.} The low energy spectrum of UMSSM has
an extra CP-even Higgs boson and two more neutralinos.  
We have shown that the SM-like Higgs boson is most of the time
the second lightest Higgs boson while the lightest one is more
singlet-like.  The singlet-like neutralino can be substantially
lighter than the MSSM lightest neutralino, such that the
SM-like Higgs boson can decay invisibly more frequently once this mode is open,
but not quite so in the MSSM.

{\it Comparison with the NMSSM.} In terms of particle content, the major
differences between UMSSM and NMSSM include 
\begin{itemize}
\item
NMSSM has two pseudoscalar
Higgs bosons while UMSSM only has one, because the would-be-Goldstone 
boson becomes the longitudinal component of the $Z'$ boson.
\item 
NMSSM has five neutralinos with the extra one coming from the
singlino while UMSSM has six neutralinos with additional ones
from the singlino and $Z'$-ino.
\item 
UMSSM also has a $Z'$ boson at TeV scale. 
\end{itemize}
\noindent
Due to the first difference the SM-like Higgs boson in NMSSM often 
decays into two light pseudoscalar Higgs boson \cite{gunion}. 
If so the production rates into $\gamma\gamma$, $WW$, $ZZ$, and
$b\bar b$ would be substantially smaller than the current data. 
However, some parameter-space points are uncovered recently such that
the SM-like Higgs boson can decay very similarly to the SM 
Higgs boson \cite{jack,ell-higgs}.

To recap, the search for the final missing piece of the SM, the Higgs
boson, remains a tantalizing task for both experimentalists and
theorists.  We have demonstrated that in models beyond the SM like the
UMSSM, we might be entertained by a SM-like Higgs boson as a mimicker
at the LHC.  This SM-like Higgs boson can be light in the mass range
of 120 -- 130 GeV as indicated by the recent LHC data and behaves
almost the same as the SM one or it can decay dominantly into
invisible modes and therefore somewhat be hidden if it is heavier. More
data are definitely needed at the LHC for detailed studies in order to
differentiate among many variants of Higgs bosons once we go
beyond the SM.


\section*{Acknowledgments}
We thank Rong-Shyang Lu for interesting discussion on the ATLAS Higgs
searches.
This work was supported in parts by the National Science Council of
Taiwan under Grant Nos. 99-2112-M-007-005-MY3 and
98-2112-M-001-014-MY3, and the WCU program through the KOSEF
funded by the MEST (R31-2008-000-10057-0).

%

\end{document}